\documentclass[twocolumn,aps,amssymb,noshowpacs,superscriptaddress,nofootinbib]{revtex4}

\usepackage{amsmath,amssymb}
\usepackage{graphicx}
\usepackage{verbatim}
\usepackage{amsfonts}
\usepackage{dcolumn}
\usepackage{bm}
\usepackage[section]{placeins}
\usepackage{color}
\usepackage{mathrsfs}
\usepackage[colorlinks,bookmarks=true,citecolor=blue,linkcolor=blue,urlcolor=red]{hyperref}
\usepackage{afterpage}
\usepackage{ulem}

\usepackage{longtable}
\usepackage{multirow}

\newcommand{\beq}{\begin{eqnarray} }
\newcommand{\eeq}{\end{eqnarray} }
\newcommand{\Beq}{\begin{eqnarray*} }
\newcommand{\Eeq}{\end{eqnarray*} }

\newcommand{\RNum}[1]{\uppercase\expandafter{\romannumeral #1\relax}}

\renewcommand{\Re}{\operatorname{Re}}
\renewcommand{\Im}{\operatorname{Im}}
\newcommand{\R}{\color{red}}
\newcommand{\B}{\color{blue}}
\newcommand{\G}{\color{green}}
\newcommand{\Y}{\color{yellow}}

\newcommand{\del}[1]{ {\R\sout{#1}}}
\newcommand{\add}[1]{{\G #1}}
\newcommand{\rep}[2]{{ \R\sout{#1}} {\B#2}}

\begin{document}
\draft

\title{Symmetry-protected Nodal Points and Nodal Lines in Magnetic Materials}

\author{Jian Yang} 
\affiliation{Beijing National Laboratory for Condensed Matter Physics, and Institute
of Physics, Chinese Academy of Sciences, Beijing 100190, China}

\author{Chen Fang}
\affiliation{Beijing National Laboratory for Condensed Matter Physics, and Institute
of Physics, Chinese Academy of Sciences, Beijing 100190, China}
\affiliation{Songshan Lake Materials Laboratory , Dongguan, Guangdong 523808, China}
\affiliation{Kavli Institute for Theoretical Sciences, Chinese Academy of Sciences, Beijing 100190, China}

\author{Zheng-Xin Liu}
\email{liuzxphys@ruc.edu.cn}
\affiliation{Department of physics, Renmin University, Beijing 100876, China}
\affiliation{Tsung-Dao Lee Institute \& School of Physics and Astronomy, Shanghai Jiao Tong University, Shanghai 200240, China}

\date{\today}
\begin{abstract}

Nodal-point  and Nodal-line structures in the dispersion of electron energy bands are characterized by their high degeneracy in certain corners or lines in the Brillouin zone (BZ). These nodal structures can also exist in the dispersion of itinerant electrons in magnetically ordered materials whose symmetry groups are anti-unitary groups called the magnetic space groups. In the present work, we provide a complete list of magnetic space groups which can host symmetry-protected nodal-point/line band structures for spin-1/2 fermionic particles, where the degeneracies at the nodal points/lines are guaranteed by irreducible projective representations (IPReps) of the little co-groups. Our discussion is restricted to the magnetic space groups whose magnetic point group contains the space-time inversion operation $\tilde T=\mathcal IT$, the combined operation of spacial inversion $\mathcal I$ and time reversal $T$, such that the energy bands are at least doubly-degenerate at arbitrary points in the BZ. For these magnetic point groups we provide the invariants to label the classes of projective Reps, and for each class we calculate all the inequivalent IPReps. From the results we select out all the groups and the corresponding Rep classes which support high-dimensional ($d\geq$4) IPReps. We then list the magnetic space groups and their high symmetry points/lines whose little co-groups have high-dimensional ($d\geq$4) IPReps with the corresponding factor systems. Examples of candidate materials are discussed.

\end{abstract}

\maketitle

\section{Introduction}\label{intro}

Gapped topological phases of matter such as topological insulators\cite{xlsc11,mzkane10,ravjzaa13} or topological superconductors\cite{ReadGreen2000,xlsc09} have nontrival band structures and novel transport properties. Some gapless materials called the topological semimetals also exhibit anomalous transport phenomena such as the negative magnetic resistance effect \cite{negmagresist16,negmagresist20} and the so-called planar Hall effect\cite{planahallef17}. Typical examples include the well studied Dirac semimetals and Weyl semimetals, whose low-energy physical properties are similar to the Dirac fermions and Weyl fermions in high energy physics, respectively. Furthermore, some low-energy quasiparticles in condensed matter have no counterpart in high energy physics, such as the semimetals with nodal lines or quadratic nodal points which have also attracted lots of interests.
For the systems whose magnetic point group contains the combined operation $\mathcal IT$ of spatial inversion $\mathcal{I}$ and time reversal $T$, namely, if $\{\mathcal IT |\tau_{\mathcal T}\}$ (with $\tau_{\mathcal T}$ a zero or nonzero fractional translation) is a symmetry element of the magnetic space group,  then the energy band has Kramers degeneracy in the whole Brillouin zone(BZ) with the degrees of the degeneracy $d\geq 2$. In this case, the nodal point/line structures are characterized by high degrees of degeneracy with $d\geq$4.
For instance, the 2 dimensional Dirac semimetal\cite{neto09} was found in honeycomb lattice such as the graphene, where the 4-fold degeneracy of the nodal point is protected by spin-rotation symmetry and the $D_{3d}$ point group symmetry.
In 3-dimensional spin-orbital coupled materials, the 4-fold degeneracy of the Dirac cones at the high symmetry points (HSPs) are protected by IPReps of the little co-groups. Especially, for certain non-symmorphic space groups, the 4-fold degeneracy is guaranteed since the lowest dimension of the IPReps of the little co-group is four\cite{CLKane2012}.

While the topological semimetals in non-magentic materials have been profoundly studied\cite{CLKane2012,burkprl11,burkprb11, BJYNagaosa14, ZFXD14,PD14,weng15,cf15,cf16, watan16, Benervig16Sci, bernevig17, watan17,kruth17,cf181,cf182,khalaf18,armitage18, cf19,bernevig19, xgw19, cano21, Bi3exp_21, Bi3thr_21}, nodal-point/line band structures for the itinerant electrons in magnetically ordered systems are less known\cite{gangxu16, gangxu18, watan18,mfvjur19,cano19, bernevig20,bernevigmtqc20,rjslager20}. Besides the difference in the intensities of
electronic correlation interactions,
the magnetically ordered materials mainly differ with the non-magnetic ones by their symmetry groups. For the non-magnetic crystals, their spacial symmetries are described by the 230 space groups, also called the type-I Shubnikov magnetic space groups. (Since time reversal is generally a symmetry, the complete symmetry group of a non-magnetic crystal is the direct product of a space group and the time reversal group, called the type-II Shubnikov magnetic space group). For magnetic materials, the time reversal symmetry is explicitly broken but the combination of time reversal and certain rotation or fractional translation remains to be a symmetry. In this case, the complete symmetry group is either a type-III or a type-IV Shubnikov magnetic space group. Owing to the rich structure of the anti-unitary groups, the type-III and type-IV magnetic space groups can yield rich band structures for the itinerant electrons in magnetic materials.

In the present work, we study symmetry-protected nodal-point and nodal-line structures for spin-1/2 fermionic particles in magnetic semimetals, where the multipole degeneracies are guaranteed by IPReps of the little co-groups. We only consider the case where the energy bands are always at least doubly degenerate away from the high symmetry points/lines, this requires that the combined operation $\tilde T$ is an element of the little co-group of magnetic space group (this means that $\{\tilde T | \pmb \tau_{\tilde T}\}$ is an element of the magnetic space group, where $\pmb \tau_{\tilde T}$ is either a fractional translation or a zero vector). By calculating the IPReps of the little co-groups, we provide a complete list of magnetic space groups (which contain $\{\tilde T|\pmb \tau_{\tilde T}\}$ as a symmetry element) and the high symmetry points/lines which support the high degeneracy ($d\geq 4$). We also provide the dispersion around the nodal points/lines. These results provide important information for experimental realization of nodal-point/line semimetals in magnetic materials.

The paper is organized as follows. In section \ref{sec: IPRep}, we provide the IPReps of the magnetic point groups which contain the $\tilde T=\mathcal IT$ operation. We considered all classes of projective Reps which are classified by the second group cohomology. For each classes, we calculate all the inequivalent IPReps. In section \ref{sec:disp}, we provide the criterion to judge the dispersion around the high-degeneracy points/lines, from which we can judge if the dispersion around a nodal point is a Dirac cone or a quadratic band touch, or a combination of them. The conditions for the existence of nodal-line band structures are also provided.  Section \ref{sec:tabinstruct} is an instruction to read the information listed in the tables, where the central results of the present work are summarized in Tab.~\ref{tab: NodalDisp}. Some potential materials are proposed in section \ref{sec:mats}. Section \ref{sec: sum} is devoted to the conclusions and discussions.

\section{spectrum degeneracy: IPReps of the little co-groups}\label{sec: IPRep}
For a finite group $G$, the group element $g$ is represented by $M(g)$ if $g$ is a unitary element and represented by $M(g)K$ if $g$  is anti-unitary, where $M(g)$ is a unitary matrix, $K$ is the complex-conjugate operator satisfying $KU=U^*K$ with $U$ an arbitrary matrix and $U^*$ its complex conjugation.
The projective Rep of $G$ is defined as \cite{MorHam}
\begin{equation}\label{projrep}
M(g_1)K_{s(g_1)}M(g_2)K_{s(g_2)} =\omega_{2}(g_1,g_2) M(g_1g_2)K_{s(g_1g_2)},
\\
\end{equation}
for  $g_1,g_2\in G$, where $s(g)=-1, K_{s(g)}=K$ if $g$ is anti-unitary and $s(g)=1, K_{s(g)}=I$ (identity matrix) if $g$ is unitary, $\omega_{2}(g_1,g_2)$ is the factor system of projective Reps with $|\omega_{2}(g_1,g_2)|=1$. If $\omega_{2}(g_1,g_2)=1$ for any $g_{1},g_{2}\in G$, above projective Rep is trivial, namely, it is a linear Rep. Under a gauge transformation $M'(g)K_{s(g)}=M(g)\Omega_1(g)K_{s(g)}$ with $|\Omega_1(g)|=1$, the Eq.\eqref{projrep} changes into
\begin{equation}\label{}
M'(g_1)K_{s(g_1)}M'(g_2)K_{s(g_2)} =\omega'_{2}(g_1,g_2) M'(g_1g_2)K_{s(g_1g_2)}
\\
\end{equation}
with the factor system
\beq\label{newfactsys}
\omega'_2(g_1,g_2)=\omega_2(g_1,g_2)\frac{\Omega_1(g_1)\Omega_1^{s(g_1)}(g_2)}{\Omega_1(g_1g_2)}.
\eeq
The two factor systems $\omega_{2}$ and $\omega'_{2}$ belong to the same class.
We only use one factor system $\omega_{2}(g_1,g_2)$ of each class to construct the regular projective Rep of $G$
(see appendix \ref{regularProj}).
The IPReps of $G$ can be obtained by reducing the regular projective Rep of $G$.

As mentioned, we consider the type-\textrm{III} or type-\textrm{IV} Shubnikov space groups as the symmetry groups of magnetic materials.
 For the type-\textrm{III} Shubnikov space groups\cite{BradleyCrack}, every anti-unitary symmetry element $T\{g|\pmb \tau_{g}\}$ is a combination of time reversal $T$ and certain space group operation with fractional translation $\pmb \tau_{g}$ in which the point group operation $g\neq E$.
Therefore this type of groups have the structure $\mathcal{M}=\mathcal{H}+T(G-\mathcal{H})$, where $\mathcal{H}$ is a halving subgroup of Fedorov (ordinary) space group $G$. In this case, time reversal operation $T$ is not an element of $\mathcal{M}$.
On the other hand, in a type \textrm{IV} Shubnikov space group \cite{BradleyCrack} the combination of time reversal $T$ and certain fractional translation $\pmb \tau_0$ is a symmetry operation. Therefore this type of groups have the structure $\mathcal M=G+T\{E|\pmb \tau_0\}G$, where $G$ is a Fedorov space group.

In the Rep theory of a type-III or type-IV Shubnikov magnetic space group $\mathcal M$, the magnetic little group $\mathcal M(\pmb k)$ is the subgroup of $\mathcal M$ which transforms the wave vector $\pmb k$ in the BZ into its equivalent wave vector $\pmb k+\pmb K$, here $\pmb K$ is a reciprocal lattice vector. $\mathcal M(\pmb k)$ is also a Shubnikov space group which has translation group as its normal subgroup. The (magnetic) point group of $\mathcal M(\pmb{k})$ is called (magnetic) little co-group $G_{0}(\pmb k)$, which is the quotient group of $\mathcal M(\pmb{k})$ with respect to the translation group.

The isogonal point group of Shubnikov space group $\mathcal{M}$ is $\mathcal{M}_{0}$.
To get symmetric group $P(\pmb k)$ of $\pmb k$ at high symmetry
point or high symmetry line,
double the group elements of $P(\pmb k)$ in Table 3.6 of Ref\cite{BradleyCrack} by multiplying all the group elements by $\mathcal{I}T$.
The little co-group $G_0(\pmb k)$ is intersection of $\mathcal{M}_{0}$ and  $P(\pmb k)$, i.e.
$G_0(\pmb k)=\mathcal{M}_{0}\cap P(\pmb k)$.

A remarkable property of a linear Rep of the Shubnikov space group $\mathcal M$ is that it defines a projective Rep \eqref{projrep} of the little co-group $G_{0}(\pmb k)$ for a momentum $\pmb k$ if it is a high symmetry point (HSP) or a point on a high symmetry line (HSL) defined in the BZ.
Except for accidental degeneracy, for itinerant electrons in magnetically ordered materials, the degeneracy of the energy bands at a given momentum $\pmb k$ of HSP or HSL in the BZ is generally protected by IPReps of $G_0(\pmb k)$.

For Shubnikov magnetic space group $\mathcal M$, the fractional translations associated with the point group operations contribute to the factor system of the projective Rep of $G_{0}(\pmb k)$ as the following \cite{BradleyCrack,ChenJQRMP85,ChenJQ02},
\begin{equation}\label{phasemag}
\omega_{2b}(g_1,g_2)=e^{-i\pmb K_{1}\cdot\pmb \tau_{2} },
\end{equation}
where reciprocal lattice vector $\pmb K_{1}=s(g_1)(g_1^{-1}\pmb k -\pmb k)$, $\pmb \tau_{2}$ is fractional translation associated with $g_{2}\in G_{0}(\pmb k)$, and $s(g_1)=-1$ if $g_1$ is anti-unitary otherwise $s(g_1)=1$.
If $\mathcal M$ is symmorphic space group, the fractional translation $\pmb \tau_{g}$ is zero for any $g\in G_0(\pmb k)$, the factor system $\omega_{2b}(g_1,g_2)=1$ for any $g_1,g_2\in G_0(\pmb k)$.
Above factor system $\omega_{2b}(g_1,g_2)$ is completely determined by the group elements of $\mathcal M$. As shown in appendix \ref{proofeq2}, (\ref{phasemag}) satisfies the cocycle equation
\beq\label{2cocy2}
\omega_{2b}(g_1,g_2)\omega_{2b}(g_1g_2,g_3) = \omega_{2b}^{s(g_1)}(g_2,g_3)\omega_{2b}(g_1,g_2g_3).
\eeq
For bosonic quasi-particles (such as the magnons) with integer spin, or for fermionic quasi-particles where spin-orbital coupling is very weak, the band structure is characterized by projective Rep of $G_{0}(\pmb k)$ with the factor system $\omega_{2b}(g_1,g_2)$ provided in (\ref{phasemag}).

However, for fermionic particles with half-odd-integer spin, an extra factor system $\omega_2 ^{({1\over2})}(g_1,g_2)$ is contributed from the spin rotation owing to spin-orbit coupling. The spin-orbit coupling is always present for the itinerant electron in magnetic materials because the Zeeman coupling term
\beq
H_{\rm Zeeman} = \mu_0\sum_i \hat {\pmb S}_i\cdot {\pmb m}_i
\eeq
(where $\hat {\pmb S}_i$ is the spin operator of the fermions and ${\pmb m}_i$ is the local magnetic moment of the material) locks the point group rotations with the corresponding spin rotations and hence defines a spin-orbit coupling.

The factor system $\omega_2 ^{({1\over2})}(g_1,g_2)$ 
can be obtained from the double valued Reps of the little co-group $G_0(\pmb k)$.
The simplest double valued Rep is the one carried by spin-1/2. Any rotation operation $R_{\pmb n}(\theta)$ of angle $\theta$ along
the direction $\pmb n$ is represented as $D^{({1\over2})}(R_{\pmb n}(\theta)) = e^{-i{\theta\over2}\pmb \sigma\cdot \pmb n}$, the inversion $\mathcal I$ is represented as $D^{(1/2)}(\mathcal I)=I$, and the time reversal is represented as $i\sigma_yK$. Then $\omega_2^{(1/2)}(g_1,g_2)$ appears in the multiplication rule
\begin{eqnarray}
&&D^{({1\over2})}(g_1)K_{s(g_1)} D^{({1\over2})}(g_2)K_{s(g_2)}\nonumber\\
&=&\omega_2^{({1\over2})}(g_1,g_2)D^{({1\over2})}(g_1g_2)K_{s(g_1g_2)}.
\end{eqnarray}

Another way to obtain $\omega_2^{(1/2)} (g_1, g_2)$ is to solve the cocycle equations \eqref{2cocycle} and obtain the representative solutions for every class, then select out the right class of solution by verifying the values of the invariants (given in Table \ref{tab:invrts})
 such that they are the same as
those of the double valued Reps.
For instance, in a double valued Rep, $\chi_{S}=\frac{\omega_{2}(C_{2x},C_{2y})}{\omega_{2}(C_{2y},C_{2x})} =-1, \chi_{\tilde T}=\omega_2(\tilde T,\tilde T)=-1$ and so on.

Combining with the factor system from fractional translation, the total factor system for fermions is given by \cite{YLF20invariants}
\beq\label{fermion}
\omega_{2f}(g_1,g_2)=\omega_2^{({1\over2})}(g_1,g_2) \omega_{2b}(g_1,g_2).
\eeq

The factor systems $\omega_{2b}(g_1,g_2), \omega_{2}^{({1\over2})}(g_1,g_2), \omega_{2f}(g_1,g_2)$ generally belong to different classes of projective Reps of the little co-group $G_0(\pmb k)$, in the sense that they cannot be transformed into each other by the gauge transformation \eqref{newfactsys}.
The gauge transformations are redundant degrees of freedom because the physical properties are determined by the gauge equivalent classes and the corresponding irreducible Reps.

It turns out that all of the classes of projective Reps of $G_0(\pmb k)$ are classified by the second group cohomology $\mathcal H^2(G_{0}(\pmb k),U(1))$ and are characterized by several gauge invariants. To obtain the classification and Rep matrices of the $G_0({\pmb k})$ at the momentum $\pmb k$, we  explicitly calculate the second group cohomology $\mathcal H^2(G_{0}(\pmb k),U(1))$, provide all the gauge invariants and the corresponding factor system for each class. From every class of factor system, we construct the regular projective Rep and then obtain all the inequivalent IPReps\cite{jyzxliu2018}. The invariants are provided in Table \ref{tab:invrts} and the IPReps at the HSPs of the BZ of monoclinic, orthorhombic, tetragonal, trigonal, hexagonal and cubic lattices are given in Tabs. \ref{tab:IPReps}$\sim$\ref{tab:OhZ2T}.

From the gauge invariants, we can easily tell the classification label of the factor system given by (\ref{phasemag}) and (\ref{fermion}). As listed in Table \ref{tab: NodalDisp}, we provide the lowest dimension of IPReps of $G_{0}(\pmb k)$ at the HSPs
for fermionic particles, which is related to the degrees of degeneracy in the energy bands of electrons at these points. Next, from the Rep matrices, we can judge the dispersion around the degeneracy points/lines from $\pmb k\cdot\pmb p$ theory. The results are also listed in Table \ref{tab: NodalDisp}, and the method is discussed in the next section.

\section{Dispersion: Criteria for Dirac cones, nodal lines and others}\label{sec:disp}

Since we only consider type-III and type-IV Shubnikov magnetic space groups containing the symmetry operation $\{\tilde T|\pmb\tau_{\tilde T}\}$, the little co-groups $G_0(\pmb k)$ always contain $\tilde T$.
(Since $G_0(\pmb k)$ is the quotient group of magnetic little group $\mathcal M(\pmb{k})$ with respect to the translation group, all the elements of $G_0(\pmb k)$ have no fractional translation.)
Obviously, $\tilde T$ commutes with all the other group elements in $G_{0}(\pmb k)$, therefore $G_{0}(\pmb k)$ has the following structure
\beq\label{G0k}
G_0(\pmb k) = H\times Z_2^{\tilde T}
\eeq
with $H$ a unitary point group and $Z_2^{\tilde T}=\{E, \tilde T\}$.

We assume that the little co-group $G_0(\pmb k)$ at momentum $\pmb k$ supports high-dimensional($d\geq4$) IPReps whose Rep matrices are known. In this section, we provide the method to judge if the dispersion around the point $\pmb k$ is linear or quadratic, and if the degeneracy is stable in a HSL. Thus we can know the symmetry groups that can host symmetry-protected Dirac cones, quadratic band touching nodal points, nodal lines and other dispersion relations.

\subsection{Nodal points with linear or higher-order dispersions }\label{nodalpoint}

Supposing that at a HSP $\pmb k$ the energy eigenstates carry a $d$-dimensional ($d\geq4$) IPRep $M(G_{0}(\pmb k))$ of the little co-group $G_{0}(\pmb k)$, and that $\pmb k$ is a touching point of two energy bands. We first discuss the criteria for a Dirac-type linear dispersion.

Following the spirit of $\pmb k\cdot \pmb p$ theory, it is sufficient to consider the bands touching at the HSP $\pmb k$. We write the fermion bases of the $d$-dimensional IPRep as $|\phi_{\pmb k}^\alpha\rangle = (\psi^\alpha _{\pmb k})^\dag|{\rm vacumn}\rangle, \alpha=1,2,..., d$, then for $g\in G_{0}(\pmb k)$ we have
\beq
\hat{g} |\phi_{\pmb k}^\alpha\rangle = \sum_{\beta=1}^d |\phi_{\pmb k}^\beta \rangle M(g)_{\beta\alpha}K_{s(g)},
\eeq
or equivalently
\beq
&&\hat{g}(\psi_{\pmb k}^\dag )\hat{g}^{-1} = \psi_{\pmb k}^\dag  M(g)K_{s(g)},\label{Psik^}\\
&&\hat{g}\psi_{\pmb k}\hat{g}^{-1} = K_{s(g)} M(g)^\dag \psi_{\pmb k},\label{Psik}
\eeq
here the group element $g$ is treated as an operator $\hat{g}$ (see appendix \ref{regularProj}).
If the point $\pmb k$ is indeed a Dirac cone, then at the vicinity of $\pmb k$ the Hamiltonian should be in form of: 
\beq\label{Ham}
H_{\pmb k+\delta\pmb k} = \psi_{\pmb k+\delta\pmb k}^\dag (\delta\pmb k\cdot \pmb \Gamma) \psi_{\pmb k+\delta\pmb k},
\eeq
where $\Gamma_m,\ m=x,y,z$ are three $d\times d$ Hermitian matrices with
\beq\label{Herm}
\Gamma_m^\dag =\Gamma_m.
\eeq

When $\delta\pmb k$ is small enough, relations similar to (\ref{Psik^}) and (\ref{Psik}) hold for $\psi^\dag_{\pmb k+\delta \pmb k}$ and $\psi_{\pmb k+\delta \pmb k}$ with
\beq
&&\hat{g}(\psi_{\pmb k+\delta\pmb k}^\dag )\hat{g}^{-1} = \psi_{\pmb k + \hat g\delta\pmb k}^\dag  M(g)K_{s(g)}, \\
&&\hat{g}\psi_{\pmb k+\delta\pmb k}\hat{g}^{-1} = K_{s(g)} M(g)^\dag \psi_{\pmb k+ \hat g\delta\pmb k}.
\eeq
At the vicinity of $\pmb k$, the total Hamiltonian preserves the $G_0(\pmb k)$ symmetry, therefore,
\beq\label{HG}
\hat g_1\left(\sum_{\delta \pmb k} H_{\pmb k + \delta \pmb k}\right) \hat g_1^{-1} = \sum_{\delta \pmb k} H_{\pmb k + \delta \pmb k}
\eeq
for all $g_1\in G_0(\pmb k)$.


Now we analyze the condition under which the leading order perturbation is given by (\ref{Ham}). Notice that $\delta \pmb k$ varies as a vector under the action of $h\in H$, i.e.,
\beq
 h\delta k_m = \sum_{n} D^{(\pmb v)}_{nm}(h)\delta k_{n},\ h\in H,
\eeq
where $H$ is the
halving unitary subgroup
of $G_{0}(\pmb k)$, $D^{(\pmb v)}_{nm}(h)$ are linear combination coefficients, which are matrix elements of the vector Reps of $H$.
Moreover $\delta k_m$ is invariant under action of
$\tilde{T}=\mathcal{I}T$
\beq
&&\tilde T\delta k_m =  \delta k_m.
\eeq
Here we adopt the orthonormal bases $[\pmb b_x, \pmb b_y, \pmb b_z]$ in momentum space such that $\delta \pmb k$ has the components $(\delta k_x, \delta k_y, \delta k_z)^T$.

To ensure that the total Hamiltonian is symmetric under $G_0(\pmb k)$, $\Gamma_m$ should vary in the same way as $\delta\pmb k$ under action of $G_0(\pmb k)$,
\beq\label{vec}
&&M(h) \Gamma_m M(h)^\dag = \sum_n D^{(\pmb v)}_{nm}(h) \Gamma_n,\\
&&M(\tilde T)K \Gamma_mK M(\tilde T)^\dag = \Gamma_m.\nonumber
\eeq
The second equation is equivalent to
\beq\label{Tvec}
M(\tilde T) \Gamma_m^* M(\tilde T)^\dag = \Gamma_m.
\eeq

If there exist matrices $\Gamma_{x,y,z}$ satisfying the conditions (\ref{Herm}), (\ref{vec}) and (\ref{Tvec}), then the Hamiltonian (\ref{Ham}) yields a linear dispersion. Therefore, the existence of linear dispersion along $m$-direction is equivalent to the existence of the matrix $\Gamma_m$ satisfying above conditions.

From group theory, the condition (\ref{vec}) requires that the direct product Rep $M(h)\otimes M^*(h)$ of the halving unitary subgroup $H$ contains the vector Rep $D^{(\pmb v)}(h)$. Furthermore, the hermiticity condition (\ref{Herm}) and the $\tilde T$ condition (\ref{Tvec}) together require that the matrix $\Gamma_{m}, m=x,y,z$ satisfies the skew-symmetry condition $\tilde \Gamma_{m}^T = -\tilde \Gamma_m$ with $\tilde \Gamma_m=\Gamma_{m} M^T(\tilde T)$ (see \cite{YangYFL_CG} for detailed derivation). All of the above conditions can be checked by calculating a single quantity
{\small
\begin{eqnarray}\label{ak}
a_{\pmb k}\! \! &=&\! \!{1\over 2|H|} \sum_{h\in H} \left[ |\chi(h)|^2 \!  +\!  \omega_2(\tilde Th, \tilde Th) \chi((\tilde Th)^2)\right]\! \chi^{(\pmb v)\ast}\! (h)\notag\\
\! \!&=&\! \!{1\over 2|H|} \sum_{h\in H} \Big[ |\chi(h)|^2 \!  +\!  {\rm Tr} [M(\tilde Th)M^*(\tilde Th)]\Big]\chi^{(\pmb v)\ast}\! (h),
\end{eqnarray}
}where $a_{\pmb k}$ is the number of independent set of matrices $\Gamma_{x,y,z}$ satisfying the conditions \eqref{Herm}, \eqref{vec}, \eqref{Tvec},
and $\chi^{(\pmb v)}(h)= {\rm Tr} [D^{(\pmb v)}(h)]$ is the character of vector Rep of $D^{(\pmb v)}(h),h\in H$.
If $a_{\pmb k}$ is a nonzero integer, then the matrices $\Gamma_{x,y,z}$ can be found and the dispersion is linear along all directions, otherwise the dispersion maybe quadratic.

Above we have assumed that the vector Rep $D^{(\pmb v)}(H)$ of $H$ is irreducible and hence $\Gamma_{x,y,z}$ can be transformed into each other by the action of the little co-group $G_{0}(\pmb k)$. For most point groups, $D^{(\pmb v)}(H)$ is reducible (unless $H$ is one of the high symmetry groups $T, T_h, T_d, O, O_h$). In that case, linear dispersion may only exist along special directions and we need to check them separately. Suppose a linear Rep $(\mu)$ is included in the vector Rep, then the linear dispersion along the directions of the corresponding bases can be judged from
\beq\label{amu}
a_{\mu=}{1\over 2|H|} \sum_{h\in H} \Big[ |\chi(h)|^2 \!  +\!  {\rm Tr} [M(\tilde Th)M^*(\tilde Th)]\Big]\chi^{(\mu)\ast}\! (h).
\eeq

For instance, the vector Rep of $H=D_{4h}$ is reduced into $A_{2u}\oplus E_u$, and accordingly the vector $\delta \pmb k$ is separated into $(k_z)$ and $(k_x, k_y)^T$. Therefore, we can judge the linear dispersion along $k_z$ direction by calculating
\begin{equation}
a_{z}\! =\! {1\over 2|H|}\! \sum_{h\in H}\! \left[ |\chi(h)|^2 \! +\!   {\rm Tr\ } [M(\tilde Th)M^*(\tilde Th)] \right]\chi^{(A_{2u})\ast}(h),
\\
\end{equation}
and judge the linear dispersion along $k_x, k_y$ directions by calculating
\begin{equation}
a_{xy}\! =\!{1\over 2|H|}\! \sum_{h\in H}\! \left[ |\chi(h)|^2 \! +\! {\rm Tr\ } [M(\tilde Th)M^*(\tilde Th)] \right]\chi^{(E_{u})\ast}(h).
\\
\end{equation}
Since $E_u$ is irreducible, if $a_{xy}\neq0$ then the dispersion in the $(k_x, k_y)$ plane is linear. If both $a_{xy}$ and $a_{z}$ are nonzero, then the dispersion is linear in all directions.

Above discussion can be easily generalized to judge the existence of quadratic or higher-order dispersions. As the linear dispersion terms correspond to vector Reps of halving unitary subgroup $H\subset G_{0}(\pmb k)$, quadratic or higher-order dispersion terms carry other linear Reps of $H$. The $\chi^{(\pmb v)}(h)$ in (\ref{ak})  should be replaced by character of other linear Reps of $H$.

In table \ref{tab: NodalDisp} we give the dispersion terms which are not higher than third order at the vicinity of HSP. We also provide the directions of nodal lines in the BZ. The criterion for the existence of nodal lines crossing a HSP is discussed below.

\subsection{Nodal-line dispersions}
Now we discuss possible cases where a HSL has a nodal line dispersion, namely, the band is always 4-fold degenerate along this HSL (except for some special points where the degeneracy may be higher than 4). In these cases, the little co-groups of the HSLs are $C_{nv}\times Z_2^{\tilde T}$, which are included in Table \ref{tab:invrts}.

Without loss of generality, we assume that HSL is along $k_z$ direction. Obviously $[\pmb b_z]$ carries the identity Rep of $C_{nv}$ and of $Z_2^{\tilde T}$ (and hence of $C_{nv} \times Z_2^{\tilde T}$). In  the following we start with a HSP whose little co-group is $G_{0}(\pmb k)=H\times Z_2^{\tilde T}$ with $H\supseteq C_{nv}$, and judge if the $k_z$-line crossing it forms a nodal line or not.

{\bf Case I}, the little co-group of a HSP is the same as that of the HSL crossing it, namely, $H=C_{nv}$ and $G_{0}(\pmb k)=C_{nv}\times Z_2^{\tilde T}$. Therefore, the high-dimensional IPReps at the HSP remains irreducible along the HSL. Consequently, the $k_z$ axis forms a nodal line, no matter the dispersion along $k_x, k_y$ directions is linear or not. This situation does not occur in the type-IV magnetic space groups, but indeed occurs in type-III magnetic space groups.

Since $[\pmb b_z]$ carries the identity Rep of $H$, and the product Rep $M(H)\otimes M^*(H)$ must contain the identity Rep, thus the dispersion of the degenerate energy curve at the HSP along the nodal line direction contains linear term.

{\bf Case II}, the little co-group $G_{0}(\pmb k)$ of a HSP contains $C_{nv}\times Z_2^{\tilde T}$ as a real subgroup, namley $H \supset C_{nv}$. As clarified, the HSL along the $k_z$ direction crossing this HSP has the little co-group $C_{nv}\times Z_2^{\tilde T}$. If this HSL forms a nodal line, it should satisfy one of the following two conditions.

(1) $[\pmb b_z]$ carries a nontrivial 1-dimensional Rep of $H$, and the product Rep $M(H)\otimes M^*(H)$ does not contain the Rep carried by $[\pmb b_z]$. In this case, the CG coefficient coupling to the identity Rep forms an identity matrix $\Gamma=I$, therefore, the possible perturbation along $k_z$ direction is $k_z^2I$, which does not lift the 4-fold degeneracy. In other words, the dispersion of the energy curve at the HSP along the nodal line direction is quadratic.

(2) the vector Rep of $H$ is irreducible, and $M(H)\otimes M^*(H)$ does not contain the vector Rep carried by $[\pmb b_x, \pmb b_y, \pmb b_z]$. In this case, the HSLs along $k_x, k_y, k_z$ directions are all nodal lines, and the dispersions of the energy curve at the HSP along the nodal lines are quadratic.

Another possibility is that $[\pmb b_z]$ carries the identity Rep of $H$ [and hence of $G_{0}(\pmb k)$]. In this case, $k_z$ cannot lift the degeneracy of the HSP, hence the HSL has the same degeneracy as the HSP and forms a nodal line. The product Rep $M(H)\otimes M^*(H)$ always contains the Rep carried by $[\pmb b_z]$(i.e. the identity Rep), thus the dispersion of the degenerate energy curve at the HSP along the nodal line direction is linear. However, the only point group $H$ satisfying above requirements is $C_{nv}$, which goes back to the Case-I.

In Table \ref{tab: NodalDisp}, we provide all the directions of nodal lines crossing the HSPs of the BZ.

\section{Instruction to the tables}\label{sec:tabinstruct}

The central results of the present work are listed in Tab.~\ref{tab: NodalDisp}.

The first column shows the number of magnetic space groups and the second column lists the HSPs and the corresponding little co-groups [$G_0(\pmb k)$] as well as the dimensions (${\rm dim}$) of the IPReps for fermionic particles at the given HSPs. If dim=$n$ with $n=4$ or $8$, then it means that all the IPReps are of $n$-dimensional.  For the cases with `dim=$m|n$' ($m,n$ are two integers with $m\leq n$), $m$ stands for the lowest dimension of the IPReps of the same class and $n$ is the dimension of the present IPRep.
Take the Shubinikov space group 221.97 IV as an example. The $G_0(\pmb k)$ at R point is $O_{h}\times Z_{2}^{T}$, the projective Rep of
class ($-1,+1,-1,+1,+1$) has one $8$-dim irreducible Rep and two inequivalent $4$-dim irreducible Reps (see Table \ref{tab:OhZ2T}).
For ``dim$=4|8$'', the first number 4 stands for the lowest dimension of the IPReps of $O_{h}\times Z_{2}^{T}$ and the second number 8 is the
dimension of the IPRep discussed at present (we are discussing the dispersion and nodal line the 8-dim IPRep may lead to). Similarly, ``dim$=4|4_{1,2}$'' shows that we are discussing the dispersion and nodal line the two inequivalent 4-dim IPReps may lead to.

The third column shows the values of the gauge invariants that label the second group cohomology classification of the projective Reps.


In the fourth column of Table \ref{tab: NodalDisp}, we list the lowest order dispersions, which are polynomials of $\delta \pmb k$.
 Different dispersion terms of $\delta\pmb k$ belong to different irreducible linear Reps of $G_0(\pmb k)$.
 When one irreducible linear Rep of $G_0(\pmb k)$ corresponds to linear dispersion term of $\delta\pmb k$, if it is $3$-dimensional, we denote it as $[k_{x},k_{y},k_{z}]$; if it is $2$-dimensional in $(k_{x},k_{y})$ plane, we denote it as $[k_{x},k_{y}]$; if it is $1$-dimensional only in the direction of $k_{z}$, we denote it as $k_{z}$. Here, $k_{x},k_{y},k_{z}$ are bases of irreducible linear Rep of $G_0(\pmb k)$.
When the irreducible linear Rep of $G_0(\pmb k)$ corresponds to quadratic or cubic dispersion term of $\delta\pmb k$, the bases of irreducible linear Rep can be $k_{x}k_{y}$, $k_{x}k_{z}$, $k_{y}k_{z}$, $k_{x}k_{y}k_{z}$,$[k_{z}k_{x},k_{z}k_{y}]$,
$[k_{x}k_{y},k_{x}k_{z},k_{y}k_{z}]$,$\cdots$ . The bracketed terms correspond to bases of $2$-dim or $3$-dim irreducible linear Rep of $G_0(\pmb k)$.

 The last column provides the possible nodal line directions crossing the given HSP. It should be noticed that if the little co-group belongs to one of the groups $D_{3h}\times Z_2^{\tilde T}$, $D_{6}\times Z_2^{\tilde T}$, $D_{6h}\times Z_2^{T}$, $O_{h}\times Z_2^{T}$, then there are two or three inequivalent IPReps for the given class, the information of these inequivalent IPReps is listed separately. Our results only depend on the symmetry group and are not sensitive to the details of the material.

For most of the HSPs and HSLs listed in Tab.~\ref{tab: NodalDisp}, the IPReps are of 4-dimensional (dim$=4$). We enumerate the exceptions as follows. For the group $G_0(\pmb k)=O_h\times Z_2^{T}$, some classes contain both 4-dimensional (dim=$4|4$) and 8-dimensional (dim=$4|8$) IPReps in certain classes. If the 8-dimensional IPRep has different dispersions with the 4-dimensional IPRep, then the dispersion of the former is given in a different row. Especially, for the group $G_0(\pmb k)=D_{4h}\times Z_2^T$, several classes only contain 8-dimensional IPReps (dim=$8$). When the lowest dimension of the IPReps is $4$, the dispersions of the corresponding HSP are completely determined by the terms listed in Tab.~\ref{tab: NodalDisp} and must have a nodal structure. For example, for the group 11.57 IV, the HSPs C and E both have dim=$4$, and the dispersions are linear along $k_x, k_y, k_z$ directions, so the energy dispersions at C and E must be of Dirac-cone type, no matter the energy is close to the fermi level or not.

At some HSPs, the lowest dimension of the IPReps is $2$ (marked in red color in Tab.~\ref{tab: NodalDisp}) but the same class contains 4-dimensional (dim=$2|4$) or 6-dimensional (dim=$2|6$) IPReps. For $G_0(\pmb k) =T \times Z_2 ^{\tilde T}, T_d \times Z_2 ^{\tilde T}, T_h \times Z_2 ^{T}$ with dim=$2|6$, the dispersion terms describe bands which carry the 6-dimensional IPReps. For other cases with dim$=2|4$, the dispersion terms describe the bands which carry the 4-dimensional IPReps. Different from the case with lowest dimension dim$=4$, the nodal structures are not guaranteed if dim=$2|4$ or dim=$2|6$, and the results are dependent on the details of the materials.

The information of the  intermediate steps in obtaining Tab.~\ref{tab: NodalDisp} is provided in Tabs.~\ref{tab:invrts}$\sim$\ref{tab:OhZ2T}.

Tab.\ref{tab:IPReps} shows all inequivalent IPReps and their corresponding invariants (label of classification) for anti-unitary groups (with specific factor systems) that appeared in our discussion. The set of invariants marked by $(*)$ belongs to the same class of $\omega_{2}^{({1\over2})}(g_1,g_2)$ which is the factor system of the double valued Rep (up to a gauge transformation \eqref{newfactsys}). In other words, a Rep with mark $(*)$  can be mapped to a double valued Rep under a gauge transformation followed by a unitary transformation. Since isomorphic groups have the same (projective) Reps, we consider isomorphic groups as the same abstract group when listing their IPReps in Tab.\ref{tab:IPReps}.

Here we clarify the notations that have been used. $C_{2m}, C_{4m}^{\pm}$ with $m=x,y,z$ label the 2-fold or 4-fold axis along $\hat x, \hat y, \hat z$ respectively; $C_{3j}^{\pm}$ with $j=1,2,3,4$ label four different 3-fold axes along
$\hat x + \hat y + \hat z,\hat x + \hat y - \hat z,-\hat x + \hat y + \hat z,\hat x - \hat y + \hat z$, respectively; $C_{2}$
labels 2-fold axis along $\hat z$,
$C'_{2i}$ with $i=1,2,3$ label three different 2-fold axes $\hat x,\hat x -\sqrt{3}\hat y, \hat x +\sqrt{3}\hat y$ respectively,
$C''_{2i}$ with $i=1,2,3$ label three different 2-fold axes $\hat y,\sqrt{3}\hat x+\hat y,\sqrt{3}\hat x-\hat y$ respectively;
$C_{2p}$ with $p=a,b,c,d,e,f$ label six different 2-fold axes along $\hat x +\hat y,\hat x-\hat y, \hat x +\hat z,
\hat y +\hat z, \hat x -\hat z, \hat y -\hat z$, respectively. These notations can be found in Fig 1.1, Fig 1.2 and Fig 1.3 of Ref.\cite{BradleyCrack}. Furthermore, $\mathcal{I}$ stands for spacial  inversion, $\mathcal{M}$ denotes  planar mirror reflection. Some combined operations containing the spacial inversion $\mathcal I$ are replaced by brief notations: $\mathcal IT=\tilde T$, $\mathcal{I}C_{2}=\mathcal{M}_{h}$, $\mathcal{I}C_{2m}=\mathcal{M}_{m}$, $\mathcal{I}C'_{2i}=\mathcal{M}_{di}$, $\mathcal{I}C''_{2i}=\mathcal{M}_{vi}$, $\mathcal{I}C_{2p}=\mathcal{M}_{dp}$, $\mathcal{I}C_{3}^{\pm}=S_{6}^{\mp}$, $\mathcal{I}C_{4m}^{\pm}=S_{4m}^{\mp}$, $\mathcal{I}C_{6}^{\pm}=S_{3}^{\mp}$, $\mathcal{I}C_{3j}^{\pm}=S_{6j}^{\mp}$.

All of the abstract groups under our consideration have the structure $G=H\times Z_2^{\mathbb T}$ where $H$ is a finite unitary group and $Z_2^{\mathbb T}=\{E,\mathbb T\}$ with $\mathbb T$ an abstract anti-unitary group satisfying $\mathbb T^2=E$. When mapped to the little co-groups $G_0(\pmb k)$, $H$ corresponds to a point group, and $\mathbb T$ can be interpreted as $\tilde T$ owing to (\ref{G0k}). However, in tables \ref{tab: NodalDisp} and \ref{tab:IPReps}, we identify $\mathbb T$ as $T$ if $T\in G_0({\pmb k})$
(in this case $G_0({\pmb k})$ is the little co-group of type IV magnetic space groups and $\{T|\pmb \tau_0\}$ with $\pmb \tau_0\neq 0$ is symmetry operation) 
and identify $\mathbb T$ as $\tilde T$ if $T\notin G_0({\pmb k})$. The correspondence between abstract groups and concrete magnetic point groups are listed in Tab.~\ref{tab:invrts}, where we also provide the physical meaning of the generators and the gauge invariants which label the classification of the IPReps.

The IPReps of some big groups are listed in separate tables, namely, the group $O(T_{d})\times Z_{2}^{\tilde{T}}$ is listed in Tab.~\ref{tab:OZ2T}, the group $T_{h}\times Z_{2}^{T}$ is listed in Tab.~\ref{tab:ThZ2T}, and the group $O_{h}\times Z_2^T$ is listed in Tab.~\ref{tab:OhZ2T}. Because $\mathcal{H}^2(D_{2h}\times Z_2^T,U(1))=\mathcal{H}^2(D_{4h}\times Z_2^T,U(1))=\mathcal{H}^2(D_{6h}\times Z_2^T,U(1))=\mathbb{Z}_{2}^{7}$, there are $128$ different classes of projective Reps for each of the above groups.  We do not list the matrix forms of irreducible Reps for every class.

 \section{Potential materials}\label{sec:mats}

Our symmetry rules are helpful to search for candidate magnetic materials which support Dirac-cone or nodal-line band structures . The results are consistent with the band structures for itinerant electrons obtained from first principle  calculations of a number of magnetic materials\cite{bernevig20}.

 For instance, the uranium intermetallic compounds U$_{2}$Pd$_{2}$In\cite{u2pdin94} has the 127.394 magnetic space group symmetry. A transition metal element Pd and a magnetic atom U are contained. The little co-group $G_{0}(\pmb k)$ of the M point and the A point is $D_{2d}^{2}\times Z_{2}^{\tilde{T}}$, which has a $4$-dim IPRep and supports linear dispersions in the $(k_{x},k_{y})$ plane. On the HSL V(MA) (along $k_{z}$ direction), the little co-group is $C_{2v}^{3}\times Z_{2}^{\tilde{T}}$ whose IPRep is $4$-dimensional. Therefore, U$_{2}$Pd$_{2}$In is a magnetic nodal-line semimetal.
Similar candidate nodal line materials also include
CeCo$_2$P$_2$(126.386)\cite{CeCo2P298},UP$_{2}(130.432)$\cite{p2u66},
NdZn(222.103)\cite{NdZn87}.

We list several candidate materials hosting nodal-point structures. The rare earth compound YFe$_{4}$Ge$_{2}$\cite{YFe4Ge201} has magnetic atom Fe, whose magnetic space group symmetry is 58.399. The little co-group $G_{0}(\pmb k)$ of the Z point and the S point is $D_{2}^{1}\times Z_{2}^{\tilde{T}}$ whose IPRep is $4$-dimensional. The dispersions along $k_{x},k_{y},k_{z}$ directions are all linear. So YFe$_{4}$Ge$_{2}$ hosts Dirac cone structure at the vicinity of Z and S points.  The rare earth compound LuFe$_{4}$Ge$_{2}$\cite{LuFe4Ge212} has the same magnetic group symmetry as YFe$_{4}$Ge$_{2}$, hence it is also a magnetic nodal-point semimetal. Similarly, the cubic crystal compounds NpTe, NpSe and NpS\cite{NpS95} have the magnetic group symmetry 228.139 and support nodal-point semimetal at X point.

\section{Conclusion and discussion}\label{sec: sum}
In conclusion, we study the spectrum degeneracy and the dispersion around the degenerate nodal points/lines in the BZ for itinerant electrons in magnetic materials preserving type-III or type-IV Shubnikov magnetic space group symmetries. In our discussion the degeneracies are resulting from high-dimensional IPReps of the little co-groups. We provide the criteria for judging Dirac cones, nodal lines and other type dispersions around the high degeneracy points.

For all of the type-III and type-IV Shubnikov magnetic space groups which contain $\{\tilde T|\pmb \tau_{\tilde T}\}$ as a group element (namely, $\tilde T=\mathcal IT$ belongs to the little co-group of the HSPs), we list the complete information of the HSPs and dispersion information. Our work provides guidelines for experimental realization of topological semimetals in magnetic materials.

It should be mentioned that, although our symmetry analysis predicts the existence of degeneracies at the HSPs/HSLs and the corresponding dispersions, we cannot guarantee that these nodal points/lines appear precisely at (or close) to the fermion energy. The positions of the nodal points/lines depend on the detailed chemical components of the materials.

Despite the IPReps protection, nodal points (such as the Dirac cones) can also appear at some point in certain HSLs where two bands carrying inequivalent IPReps cross each other \cite{BJYNagaosa14}. Furthermore, degenerate nodal points/lines may also exist in magnetic materials whose magnetic little co-group does not contain $\tilde T$ as an element (in this case the double degeneracy in the whole BZ is not protected).  Finally, our method can also be applied to analyze the band structure of bosonic quasiparticles (such as the magnons) or fractional excitations (such as anyons in topologically ordered phases).

\acknowledgments{We thank Z.-Y. Yang, L.-J. Zou, P.-J. Guo, D.-X. Shao, W. Ji and X.-G. Wan for valuable discussions. J.Y. and Z.-X.L. thank the support from  Ministry of Science and Technology of China (Grant No. 2016YFA0300504). J.Y. and C.F. are supported by Ministry of Science and Technology of China under grant number 2016YFA0302400, National Science Foundation of China under grant number 11674370 and
Chinese Academy of Sciences under grant number XXH13506-202 and XDB33000000. Z.-X.L. is supported by the NSF of China (Grants No.11574392 and No. 11974421), and the Fundamental Research Funds for the Central Universities and the Research Funds of Renmin University of China (Grant No. 19XNLG11).}


\begin{table*}[htbp]
\caption{Information for the IPReps for fermionic particles at the HSPs of Shubnikov magnetic space groups, where `dim' denotes the dimension of the IPReps in the given class. If dim=4 then all the IPReps of the same class are of 4-dimensional, the case dim=$8$ is similar; if dim=$4|8$ then first number 4 stands for the lowest dimension of the IPReps of the same class and the second number 8 is the dimension of the present IPRep, similar situations occur for dim=$4|4$, $2|4$, $2|6$. For a given class (namely, the same HSP), if there are more than one IPReps with the same dimension but different dispersions, then we use the subscript $_{1,2,..}$ to distinguish these Reps.  `l.c.' stands for label of classification (i.e. the values of the gauge invariants) of projective Reps of magnetic little co-groups $G_{0}(\pmb k)$, `disp.' means the dispersion at the vicinity of the given HSP (linear, quadratic or cubic $k$-terms that split the band degeneracy are provided, higher order terms are not shown), `n.l.' shows the directions of nodal lines (if any) crossing the HSP. $k_{x\pm y}$ denotes $k_x\pm k_y$, so on and so forth. } \label{tab: NodalDisp}
\centering
\begin{tabular}{ |c|ccc|c|c|c| }
\hline
Shubnikov space group&HSP& $G_{0}(\pmb k)$ &dim & l.c.& disp.&n.l.\\
\hline
11.55~\textrm{IV}
 &Z&$C_{2h}^{1}\times Z_{2}^{T}$&$4$ &  ($-1,-1,-1,-1$)&$k_{m}, k_{n}^3, k_{z}$ ($\pmb m \perp\pmb n$)  &$-$                 \\
 &C&$C_{2h}^{1}\times Z_{2}^{T}$&$4$ &  ($-1,+1,-1,-1$)&$k_{m}, k_{n}^3, k_{z}$ ($\pmb m \perp\pmb n$)   &$-$                    \\
 &D&$C_{2h}^{1}\times Z_{2}^{T}$&$4$ &  ($-1,-1,-1,-1$)&$k_{m}, k_{n}^3, k_{z}$ ($\pmb m \perp\pmb n$)    &$-$    \\
 &E&$C_{2h}^{1}\times Z_{2}^{T}$&$4$ &  ($-1,+1,-1,-1$)&$k_{m}, k_{n}^3, k_{z}$ ($\pmb m \perp\pmb n$) &$-$     \\
\hline
11.57~\textrm{IV}
 &C&$C_{2h}^{1}\times Z_{2}^{T}$&$4$ & ($-1,-1,+1,-1$)&$k_{x},k_{y},k_{z}$&$-$    \\
 &E&$C_{2h}^{1}\times Z_{2}^{T}$&$4$ & ($-1,-1,+1,-1$)&$k_{x},k_{y},k_{z}$&$-$    \\
\hline
13.70~\textrm{IV}&B&$C_{2h}^{1}\times Z_{2}^{T}$&$4$ &($-1,-1,+1,-1$)&$k_{x},k_{y},k_{z}$&$-$    \\
\end{tabular}
%
%
\hline
14.80~\textrm{IV}&B&$C_{2h}^{1}\times Z_{2}^{T}$&$4$ & ($-1,-1,+1,-1$)&$k_{x},k_{y},k_{z}$&$-$    \\
 &Z&$C_{2h}^{1}\times Z_{2}^{T}$&$4$ &  ($-1,-1,-1,-1$)&$k_{m}, k_{n}^3, k_{z}$ ($\pmb m \perp\pmb n$)      &$-$               \\
 &C&$C_{2h}^{1}\times Z_{2}^{T}$&$4$ &  ($-1,+1,-1,-1$)&$k_{m}, k_{n}^3, k_{z}$ ($\pmb m \perp\pmb n$)      &$-$                  \\
\hline
14.81~\textrm{IV}&B&$C_{2h}^{1}\times Z_{2}^{T}$&$4$ & ($-1,-1,+1,-1$)&$k_{x},k_{y},k_{z}$&$-$     \\
 &A&$C_{2h}^{1}\times Z_{2}^{T}$&$4$ & ($-1,-1,+1,-1$)&$k_{x},k_{y},k_{z}$&$-$    \\
\hline
14.82~\textrm{IV}
 &Z&$C_{2h}^{1}\times Z_{2}^{T}$&$4$ &  ($-1,-1,-1,-1$)&$k_{m}, k_{n}^3, k_{z}$ ($\pmb m \perp\pmb n$) &$-$    \\
 &C&$C_{2h}^{1}\times Z_{2}^{T}$&$4$ &  ($-1,-1,-1,-1$)&$k_{m}, k_{n}^3, k_{z}$ ($\pmb m \perp\pmb n$) &$-$    \\
\hline
\end{tabular}
\end{table*}

\begin{table*}[htbp]
(Extension of table \ref{tab: NodalDisp})\\
\begin{tabular}{ |c|ccc|c|c|c| }
\hline
Shubnikov space group&HSP& $G_{0}(\pmb k)$ &dim & l.c.&disp.&n.l.\\
\hline

48.262~\textrm{IV}&Y&$D_{2h}^{1}\times Z_{2}^{T}$&$4$ &($-1,-1,-1,+1,+1,-1,+1$)&$k_{x},k_{y},k_{z}$&$-$ \\
 &Z&$D_{2h}^{1}\times Z_{2}^{T}$&$4$ &($-1,-1,-1,+1,+1,-1,-1$) &$k_{x},k_{y},k_{z}$&$-$\\
 &U&$D_{2h}^{1}\times Z_{2}^{T}$&$4$ &($-1,+1,-1,-1,+1,-1,+1$) &$k_{x},k_{z},k_{x}k_{y}$ &$k_{y}$ \\
 &T&$D_{2h}^{1}\times Z_{2}^{T}$&$4$ &($-1,-1,-1,+1,+1,+1,-1$) &$k_{x},k_{y},k_{z}$&$-$\\
 &S&$D_{2h}^{1}\times Z_{2}^{T}$&$4$ &($-1,+1,-1,-1,+1,-1,-1$) &$k_{x},k_{y},k_{x}k_{z}$&$k_{z}$ \\
 &R&$D_{2h}^{1}\times Z_{2}^{T}$&$4$ &($-1,+1,-1,-1,+1,+1,+1$) &$k_{x},k_{x}k_{y},k_{x}k_{z},k_{y}k_{z}$ &$k_{y},k_{z}$ \\
\hline

48.263~\textrm{IV}
 &Z&$D_{2h}^{1}\times Z_{2}^{T}$&$4$ &($-1,-1,-1,+1,+1,-1,-1$)&$k_{x},k_{y},k_{z}$&$-$\\
\hline
\end{tabular}
\end{table*}

\begin{center}
\begin{table*}[htbp]
(Extension of table \ref{tab: NodalDisp})\\
\begin{tabular}{ |c|ccc|c|c|c| }
\hline
Shubnikov space group&HSP& $G_{0}(\pmb k)$ &dim & l.c.&disp.&n.l.\\
\hline

50.286~\textrm{IV}
 &X&$D_{2h}^{1}\times Z_{2}^{T}$&$4$ &($-1,-1,-1,+1,+1,+1,-1$)&$k_{x},k_{y},k_{z}$ &$-$\\
 &Z&$D_{2h}^{1}\times Z_{2}^{T}$&$4$ &($-1,+1,-1,+1,+1,+1,+1$)&$k_{z},k_{x}k_{y},k_{x}k_{z},k_{y}k_{z}$ &$k_{x},k_{y}$ \\
 &U&$D_{2h}^{1}\times Z_{2}^{T}$&$4$&($-1,+1,-1,+1,+1,+1,-1$)&$k_{y},k_{z},k_{x}k_{z}$ &$k_{x}$ \\
 &T&$D_{2h}^{1}\times Z_{2}^{T}$&$4$ &($-1,-1,-1,+1,-1,-1,+1$)&$k_{x},k_{z}$  & $k_{y}$ \\
 &S&$D_{2h}^{1}\times Z_{2}^{T}$&$4$ &($-1,+1,-1,+1,-1,-1,-1$) &$k_{x},k_{y},k_{y}k_{z}$ &$k_{z}$ \\
 &R&$D_{2h}^{1}\times Z_{2}^{T}$&$4$ &($-1,-1,-1,+1,-1,-1,-1$) &$k_{x},k_{y}$ & $k_{z}$ \\
\hline
50.287~\textrm{IV}
 &S&$D_{2h}^{1}\times Z_{2}^{T}$&$4$ &($-1,-1,-1,-1,-1,-1,-1$)&$k_{x},k_{y},k_{x}k_{y}k_{z},...$ &
 $k_{z}$ \\
\hline
\end{tabular}
\end{table*}
\end{center}

\begin{table*}[htbp]
(Extension of table \ref{tab: NodalDisp})\\
\begin{tabular}{ |c|ccc|c|c|c| }
\hline
Shubnikov space group&HSP& $G_{0}(\pmb k)$ &dim & l.c.&disp.&n.l.\\
\hline

52.315~\textrm{IV}
 &X&$D_{2h}^{1}\times Z_{2}^{T}$& $4$ &($-1,-1,-1,+1,+1,+1,-1$)&$k_{x},k_{y},k_{z}$ &$-$\\
 &Z&$D_{2h}^{1}\times Z_{2}^{T}$& $4$ &($-1,-1,-1,+1,+1,-1,-1$)&$k_{x},k_{y},k_{z}$ &$-$\\
 &U&$D_{2h}^{1}\times Z_{2}^{T}$& $4$ &($-1,-1,-1,+1,+1,-1,+1$)&$k_{x},k_{y},k_{z}$ &$-$\\
\hline
52.316~\textrm{IV}&Y&$D_{2h}^{1}\times Z_{2}^{T}$& $4$ &($+1,-1,-1,+1,-1,-1,-1$)&$k_{y},k_{z},k_{x}k_{y}$  &$k_{x}$ \\
 &X&$D_{2h}^{1}\times Z_{2}^{T}$& $4$ &($-1,-1,-1,+1,+1,+1,-1$)&$k_{x},k_{y},k_{z}$ &$-$\\
 &U&$D_{2h}^{1}\times Z_{2}^{T}$& $4$ &($-1,+1,-1,+1,+1,-1,+1$)& $k_{x},k_{z},k_{y}k_{z}$ &$k_{y}$ \\
 &S&$D_{2h}^{1}\times Z_{2}^{T}$& $4$ &($+1,-1,-1,+1,-1,-1,+1$)&$k_{y},k_{x}k_{z}$ &$k_{x},k_{z}$ \\
\hline
\end{tabular}
\end{table*}

\begin{table*}[htbp]
(Extension of table \ref{tab: NodalDisp})\\
\begin{tabular}{ |c|ccc|c|c|c| }
\hline
Shubnikov space group&HSP& $G_{0}(\pmb k)$ &dim & l.c.&disp.&n.l.\\
\hline

54.347~\textrm{IV}&Y&$D_{2h}^{1}\times Z_{2}^{T}$& $4$&($-1,+1,-1,+1,-1,+1,+1$)&$k_{y},k_{x}k_{y},k_{x}k_{z},k_{y}k_{z}$ &$k_{x},k_{z}$ \\
 &X&$D_{2h}^{1}\times Z_{2}^{T}$& $4$&($+1,-1,-1,-1,+1,-1,+1$)&$k_{x},k_{y},k_{x}k_{z}$ &$k_{z}$ \\
 &Z&$D_{2h}^{1}\times Z_{2}^{T}$& $4$&($-1,-1,-1,+1,+1,-1,-1$)&$k_{x},k_{y},k_{z} $&$-$ \\
 &U&$D_{2h}^{1}\times Z_{2}^{T}$& $4$&($+1,-1,-1,-1,+1,+1,-1$)&$k_{x},k_{y}k_{z} $&$k_{y},k_{z}$ \\
 &T&$D_{2h}^{1}\times Z_{2}^{T}$& $4$&($-1,+1,-1,+1,-1,-1,-1$)&$k_{x},k_{y},k_{y}k_{z} $&$k_{z}$ \\
 &S&$D_{2h}^{1}\times Z_{2}^{T}$& $4$&($+1,+1,-1,-1,-1,-1,+1$)&$k_{y},k_{x}k_{y},k_{x}k_{z}$ &$k_{x},k_{z}$ \\
 &R&$D_{2h}^{1}\times Z_{2}^{T}$& $4$&($+1,+1,-1,-1,-1,+1,-1$)&$k_{x},k_{x}k_{y},k_{y}k_{z} $&$k_{y},k_{z}$ \\
\hline
\end{tabular}
\end{table*}

\begin{table*}[htbp]
(Extension of table \ref{tab: NodalDisp})\\
\begin{tabular}{ |c|ccc|c|c|c| }
\hline
Shubnikov space group&HSP& $G_{0}(\pmb k)$ &dim & l.c.&disp.&n.l.\\
\hline

56.368~\textrm{III}&Y&$C_{2v}^{3}\times Z_{2}^{\tilde T}$& $4$ &($-1,-1,+1,-1$)&$k_{x},k_{y}$& $k_{z}$\\
 &X&$C_{2v}^{3}\times Z_{2}^{\tilde T}$& $4$ &($-1,-1,-1,+1$)&$k_{x},k_{y}$& $k_{z}$\\
 &U&$C_{2v}^{3}\times Z_{2}^{\tilde T}$& $4$ &($-1,-1,-1,+1$)&$k_{x},k_{y}$& $k_{z}$\\
 &T&$C_{2v}^{3}\times Z_{2}^{\tilde T}$& $4$ &($-1,-1,+1,-1$)&$k_{x},k_{y}$& $k_{z}$\\
 &S&$C_{2v}^{3}\times Z_{2}^{\tilde T}$& $4$ &($-1,-1,-1,-1$)&$k_{x},k_{y}$& $k_{z}$\\
 &R&$C_{2v}^{3}\times Z_{2}^{\tilde T}$& $4$ &($-1,-1,-1,-1$)&$k_{x},k_{y}$& $k_{z}$\\
\hline
56.371~\textrm{III}
 &Z&$D_{2}^{1}\times Z_{2}^{\tilde T}$& $4$ &($-1,-1,-1,-1$)&$k_{x},k_{y},k_{z}$&$-$ \\
\hline
\end{tabular}
\end{table*}

\begin{table*}[htbp]
(Extension of table \ref{tab: NodalDisp})\\
\begin{tabular}{ |c|ccc|c|c|c| }
\hline
Shubnikov space group&HSP& $G_{0}(\pmb k)$ &dim & l.c.&disp.&n.l.\\
\hline

57.391~\textrm{IV}
 &X&$D_{2h}^{1}\times Z_{2}^{T}$& $4$ &($-1,+1,-1,-1,+1,+1,+1$)&$k_{x},k_{x}k_{y},k_{x}k_{z},k_{y}k_{z}$ &$k_{y},k_{z}$ \\
 &Z&$D_{2h}^{1}\times Z_{2}^{T}$&$4$ &($+1,-1,-1,+1,+1,+1,-1$)&$k_{x},k_{z},k_{y}k_{z} $&$k_{y}$ \\
 &U&$D_{2h}^{1}\times Z_{2}^{T}$& $4$ &($+1,+1,-1,-1,+1,+1,-1$)&$k_{x},k_{x}k_{z},k_{y}k_{z} $&$k_{y},k_{z}$ \\
 &T&$D_{2h}^{1}\times Z_{2}^{T}$& $4$ &($-1,+1,-1,+1,+1,-1,+1$)&$k_{x},k_{z},k_{y}k_{z} $&$k_{y}$ \\
 &S&$D_{2h}^{1}\times Z_{2}^{T}$& $4$ &($+1,-1,-1,-1,+1,-1,-1$)&$k_{x},k_{z},k_{x}k_{y} $&$k_{y}$ \\
 &R&$D_{2h}^{1}\times Z_{2}^{T}$& $4$ &($-1,-1,-1,-1,+1,-1,+1$)&$k_{x},k_{z},k_{x}k_{y}k_{z},...$&
 $k_{y}$ \\
\hline
57.392~\textrm{IV}
 &X&$D_{2h}^{1}\times Z_{2}^{T}$& $4$ &($-1,+1,-1,-1,+1,+1,+1$) &$k_{x},k_{x}k_{y},k_{x}k_{z},k_{y}k_{z}$ &$k_{y},k_{z}$ \\
\hline
\end{tabular}
\end{table*}

\begin{table*}[htbp]
(Extension of table \ref{tab: NodalDisp})\\
\begin{tabular}{ |c|ccc|c|c|c| }
\hline
Shubnikov space group&HSP& $G_{0}(\pmb k)$ &dim & l.c.&disp.&n.l.\\
\hline

60.420~\textrm{III}
 &Z&$C_{2v}^{2}\times Z_{2}^{\tilde T}$& $4$&($-1,-1,+1,-1$) &$k_{x},k_{z}$& $k_{y}$\\
 &T&$C_{2v}^{2}\times Z_{2}^{\tilde T}$& $4$&($-1,-1,+1,-1$) &$k_{x},k_{z}$& $k_{y}$\\
\hline
60.421~\textrm{III}
 &X&$C_{2v}^{3}\times Z_{2}^{\tilde T}$& $4$&($-1,-1,-1,+1$) &$k_{x},k_{y}$& $k_{z}$\\
 &U&$C_{2v}^{3}\times Z_{2}^{\tilde T}$& $4$&($-1,-1,-1,+1$) &$k_{x},k_{y}$& $k_{z}$\\
\hline
60.425~\textrm{III}&Y&$D_{2}^{1}\times Z_{2}^{\tilde T}$& $4$&($-1,-1,-1,+1$)&$k_{x},k_{y},k_{z}$& $-$\\
 &U&$D_{2}^{1}\times Z_{2}^{\tilde T}$& $4$&($-1,-1,+1,-1$)&$k_{x},k_{y},k_{z}$& $-$\\
 &R&$D_{2}^{1}\times Z_{2}^{\tilde T}$& $4$&($-1,-1,-1,-1$) &$k_{x},k_{y},k_{z}$& $-$\\
\hline
\end{tabular}
\end{table*}

\begin{table*}[htbp]
(Extension of table \ref{tab: NodalDisp})\\
\begin{tabular}{ |c|ccc|c|c|c| }
\hline
Shubnikov space group&HSP& $G_{0}(\pmb k)$ &dim & l.c.&disp.&n.l.\\
\hline

62.450~\textrm{IV}&Y&$D_{2h}^{1}\times Z_{2}^{T}$&$4$ &($+1,-1,-1,+1,-1,-1,-1$) &$k_{y},k_{z},k_{x}k_{y} $&$k_{x}$ \\
 &Z&$D_{2h}^{1}\times Z_{2}^{T}$& $4$ &($+1,-1,-1,+1,+1,-1,+1$)&$k_{y},k_{z},k_{x}k_{z} $&$k_{x}$ \\
 &U&$D_{2h}^{1}\times Z_{2}^{T}$& $4$ &($-1,+1,-1,+1,+1,+1,+1$) &$k_{z},k_{x}k_{y},k_{x}k_{z},k_{y}k_{z} $&$k_{x},k_{y}$ \\
 &T&$D_{2h}^{1}\times Z_{2}^{T}$& $4$ &($-1,-1,-1,+1,-1,+1,-1$) &$k_{y},k_{z},k_{x}k_{y}k_{z},... $&
 $k_{x}$ \\
 &S&$D_{2h}^{1}\times Z_{2}^{T}$& $4$ &($-1,+1,-1,+1,-1,+1,-1$) &$k_{y},k_{z},k_{x}k_{y} $&$k_{x}$ \\
 &R&$D_{2h}^{1}\times Z_{2}^{T}$& $4$ &($+1,+1,-1,+1,-1,-1,-1$) &$k_{z},k_{x}k_{y},k_{y}k_{z} $&$k_{x},k_{y}$ \\
\hline
62.451~\textrm{IV}
 &X&$D_{2h}^{1}\times Z_{2}^{T}$& $4$ &($+1,-1,-1,-1,+1,-1,+1$) &$k_{x},k_{y},k_{x}k_{z}$ &$k_{z}$ \\
\hline
\end{tabular}
\end{table*}

\begin{table*}[htbp]
(Extension of table \ref{tab: NodalDisp})\\
\begin{tabular}{ |c|ccc|c|c|c| }
\hline
Shubnikov space group&HSP& $G_{0}(\pmb k)$ &dim & l.c.&disp.&n.l.\\
\hline

68.519~\textrm{IV}
 &Z&$D_{2h}^{1}\times Z_{2}^{T}$& $4$ &($-1,-1,-1,+1,+1,-1,-1$)&$k_{x},k_{y},k_{z} $& $-$\\
 &T&$D_{2h}^{1}\times Z_{2}^{T}$& $4$ &($-1,-1,-1,+1,+1,-1,-1$) &$k_{x},k_{y},k_{z} $&$-$ \\
\hline
68.520~\textrm{IV}
 &R&$C_{2h}^{1}\times Z_{2}^{T}$& $4$ &($-1,-1,-1,-1$)&$k_{m}, k_{n}^3, k_{z}$ ($\pmb m \perp\pmb n$) & $-$\\
\hline
70.531~\textrm{III}&Y&$D_{2}^{1}\times Z_{2}^{\tilde T}$& $4$&($-1,-1,-1,+1$)&$k_{x},k_{y},k_{z}$& $-$\\
&X&$D_{2}^{1}\times Z_{2}^{\tilde T}$& $4$&($-1,-1,+1,-1$) &$k_{x},k_{y},k_{z}$& $-$\\
&Z&$D_{2}^{1}\times Z_{2}^{\tilde T}$& $4$&($-1,-1,-1,-1$) &$k_{x},k_{y},k_{z}$& $-$\\
\hline
70.532~\textrm{IV}&Y&$D_{2h}^{1}\times Z_{2}^{T}$&$4$ &($-1,-1,-1,+1,+1,-1,+1$) &$k_{x},k_{y},k_{z} $&$-$ \\
\hline
\end{tabular}
\end{table*}

\begin{table*}[htbp]
(Extension of table \ref{tab: NodalDisp})\\
\begin{tabular}{ |c|ccc|c|c|c| }
\hline
Shubnikov space group&HSP& $G_{0}(\pmb k)$ &dim & l.c.&disp.&n.l.\\
\hline

124.356~\textrm{III}
 &Z&$D_{2d}^{1}\times Z_{2}^{\tilde T}$& $4$ &($-1,-1,-1,+1$)&$[k_{x},k_{y}],k_{z}$& $-$\\
 &A&$D_{2d}^{1}\times Z_{2}^{\tilde T}$& $4$ &($-1,-1,-1,+1$) &$[k_{x},k_{y}],k_{z}$& $-$\\
 &R&$D_{2}^{1}\times Z_{2}^{\tilde T}$& $4$ &($-1,-1,-1,-1$) &$k_{x},k_{y},k_{z}$& $-$\\
\hline
124.358~\textrm{III}
 &Z&$D_{2d}^{2}\times Z_{2}^{\tilde T}$& $4$ &($-1,-1,+1,-1$)&$[k_{x},k_{y}],k_{z}$& $-$\\
 &A&$D_{2d}^{2}\times Z_{2}^{\tilde T}$&$4$ &($-1,-1,+1,-1$) &$[k_{x},k_{y}],k_{z}$& $-$\\
\hline
124.359~\textrm{III}
 &Z&$D_{4}^{1}\times Z_{2}^{\tilde T}$& $4$ &($-1,-1,-1,-1$) &$[k_{x},k_{y}],k_{z}$& $-$\\
 &A&$D_{4}^{1}\times Z_{2}^{\tilde T}$& $4$ &($-1,-1,-1,-1$) &$[k_{x},k_{y}],k_{z}$& $-$\\
\hline
\end{tabular}
\end{table*}

\begin{table*}
(Extension of table \ref{tab: NodalDisp})\\
\begin{tabular}{ |c|ccc|c|c|c| }
\hline
Shubnikov space group&HSP& $G_{0}(\pmb k)$ &dim & l.c.&disp.&n.l.\\
\hline

127.392~\textrm{III}&M&$D_{2d}^{1}\times Z_{2}^{\tilde T}$& $4$&($-1,-1,-1,+1$) &$[k_{x},k_{y}],k_{z}$& $-$\\
 &A&$D_{2d}^{1}\times Z_{2}^{\tilde T}$& $4$&($-1,-1,-1,+1$)&$[k_{x},k_{y}],k_{z}$& $-$\\
\hline

127.394~\textrm{III}&M&$D_{2d}^{2}\times Z_{2}^{\tilde{T}}$& $4$&($-1,-1,-1,+1$) &$[k_{x},k_{y}]$& $k_{z}$\\
 &A&$D_{2d}^{2}\times Z_{2}^{\tilde{T}}$& $4$&($-1,-1,-1,+1$)&$[k_{x},k_{y}]$& $k_{z}$\\
\hline
127.395~\textrm{III}&M&$D_{4}^{1}\times Z_{2}^{\tilde T}$& $4$&($-1,-1,-1,+1$)&$[k_{x},k_{y}],k_{z}$& $-$\\
 &A&$D_{4}^{1}\times Z_{2}^{\tilde T}$& $4$&($-1,-1,-1,+1$) &$[k_{x},k_{y}],k_{z}$& $-$\\
\hline
127.396~\textrm{IV}&M&$D_{4h}^{1}\times Z_{2}^{T}$&$4$ &($-1,-1,-1,-1,+1,+1,+1$)&
$k_{x}k_{y},[k_{z}k_{x},k_{z}k_{y}]$& $k_{x},k_{y},k_{z}$\\
\hline
\end{tabular}
\end{table*}

\begin{table*}
(Extension of table \ref{tab: NodalDisp})\\
\begin{tabular}{ |c|ccc|c|c|c| }
\hline
Shubnikov space group&HSP& $G_{0}(\pmb k)$ &dim & l.c.&disp.&n.l.\\
\hline

130.431~\textrm{III}
 &Z&$D_{4}^{1}\times Z_{2}^{\tilde T}$& $4$&($-1,-1,-1,-1$)  &$[k_{x},k_{y}],k_{z}$& $-$\\
 &A&$D_{4}^{1}\times Z_{2}^{\tilde T}$& $4$&($-1,-1,-1,-1$)  &$[k_{x},k_{y}],k_{z}$& $-$\\
\hline
130.432~\textrm{IV}&M&$D_{4h}^{1}\times Z_{2}^{T}$&$4$ &($-1,-1,-1,-1,+1,-1,+1$) &
$[k_{x},k_{y}],k_{x}k_{y}k_{z},... $& $k_{z}$\\
 &A&$D_{4h}^{1}\times Z_{2}^{T}$& $4$ &($-1,+1,-1,-1,+1,+1,-1$) &
 $k_{x}k_{y},[k_{z}k_{x},k_{z}k_{y}]$& $k_{x},k_{y},k_{z}$\\
 &R&$D_{2h}^{1}\times Z_{2}^{T}$& $4$ &($+1,+1,-1,+1,-1,-1,+1$) &
 $k_{y},k_{x}k_{z},k_{y}k_{z} $& $k_{x},k_{z}$\\
 &X&$D_{2h}^{1}\times Z_{2}^{T}$& $4$ &($+1,-1,-1,+1,-1,+1,-1$) &$k_{x},k_{y},k_{y}k_{z} $& $k_{z}$\\
\hline
130.433~\textrm{IV}&M&$D_{4h}^{1}\times Z_{2}^{T}$& $4$ &($-1,-1,-1,+1,+1,-1,+1$)&$[k_{x},k_{y}],k_{z} $& $-$\\
\hline
\end{tabular}
\end{table*}

\begin{table*}
(Extension of table \ref{tab: NodalDisp})\\
\begin{tabular}{ |c|ccc|c|c|c| }
\hline
Shubnikov space group&HSP& $G_{0}(\pmb k)$ &dim & l.c.&disp.&n.l.\\
\hline

133.468~\textrm{IV}&M&$D_{4h}^{1}\times Z_{2}^{T}$&$4$ &($-1,-1,-1,+1,+1,-1,+1$) &$[k_{x},k_{y}],k_{z} $& $-$\\
 &Z&$D_{4h}^{1}\times Z_{2}^{T}$& $4$ &($-1,+1,-1,+1,+1,+1,-1$) &
 $k_{z},k_{x}k_{y},[k_{z}k_{x},k_{z}k_{y}] $& $k_{x},k_{y}$\\
 &R&$D_{2h}^{1}\times Z_{2}^{T}$& $4$ &($-1,+1,-1,+1,+1,-1,+1$) &$k_{x},k_{z},k_{y}k_{z} $& $k_{y}$\\
 &X&$D_{2h}^{1}\times Z_{2}^{T}$& $4$ &($-1,-1,-1,+1,+1,-1,+1$) &$k_{x},k_{y},k_{z} $& $-$\\
\hline

133.469~\textrm{IV}&M&$D_{4h}^{1}\times Z_{2}^{T}$& $4$ &($-1,-1,-1,-1,+1,-1,+1$) &
$[k_{x},k_{y}],k_{x}k_{y}k_{z},... $& $k_{z}$\\
 &Z&$D_{4h}^{1}\times Z_{2}^{T}$& $4$ &($-1,-1,-1,+1,+1,+1,-1$) &$[k_{x},k_{y}],k_{z} $& $-$\\
 &A&$D_{4h}^{1}\times Z_{2}^{T}$& $4$ &($-1,-1,-1,-1,+1,-1,-1$) &
 $[k_{x},k_{y}],k_{x}k_{y}k_{z},... $& $k_{z}$\\
\hline
\end{tabular}
\end{table*}

\begin{table*}
(Extension of table \ref{tab: NodalDisp})\\
\begin{tabular}{ |c|ccc|c|c|c| }
\hline
Shubnikov space group&HSP& $G_{0}(\pmb k)$ &dim & l.c.&disp.&n.l.\\
\hline

136.504~\textrm{IV}&M&$D_{4h}^{1}\times Z_{2}^{T}$&$4$ &($-1,-1,-1,-1,+1,+1,+1$) &
$k_{x}k_{y},[k_{z}k_{x},k_{z}k_{y}]$& $k_{x},k_{y},k_{z}$\\
 &Z&$D_{4h}^{1}\times Z_{2}^{T}$& $4$ &($-1,+1,-1,+1,+1,-1,+1$) &
 $k_{z},k_{x}^{2}-k_{y}^{2},[k_{z}k_{x},k_{z}k_{y}] $& $k_{x\pm y}$\\
 &A&$D_{4h}^{1}\times Z_{2}^{T}$& $8$ &($-1,+1,-1,-1,+1,-1,+1$) &
 $[k_{x},k_{y}],k_{z} $& $-$\\
 &X&$D_{2h}^{1}\times Z_{2}^{T}$& $4$ &($+1,-1,-1,+1,-1,-1,-1$) &$k_{y},k_{z},k_{x}k_{y} $& $k_{x}$\\
\hline
136.505~\textrm{IV}
 &Z&$D_{4h}^{1}\times Z_{2}^{T}$& $4$ &($-1,-1,-1,+1,+1,-1,+1$)&$[k_{x},k_{y}],k_{z} $&  $-$\\
 &A&$D_{4h}^{1}\times Z_{2}^{T}$& $4$ &($-1,-1,-1,+1,+1,-1,+1$)&$[k_{x},k_{y}],k_{z} $&  $-$\\
\hline
136.506~\textrm{IV}
 &Z&$D_{4h}^{1}\times Z_{2}^{T}$& $4$ &($-1,+1,-1,+1,+1,-1,+1$) &
 $k_{z},k_{x}^{2}-k_{y}^{2},[k_{z}k_{x},k_{z}k_{y}] $& $k_{x\pm y}$\\
\hline
\end{tabular}
\end{table*}

\begin{table*}[htbp]
(Extension of table \ref{tab: NodalDisp})\\
\centering
\begin{tabular}{ |c|ccc|c|c|c| }
\hline
Shubnikov s.g.&HSP& $G_{0}(\pmb k)$ &dim & l.c.&disp.&n.l.\\
\hline

141.560~\textrm{IV}&Z&$D_{4h}^{1}\times Z_{2}^{T}$&$4$ &($-1,-1,-1,+1,+1,+1,-1$) &$[k_{x},k_{y}],k_{z} $& $-$\\
&X&$D_{2h}^{2}\times Z_{2}^{T}$& $4$ &($-1,-1,-1,+1,+1,-1,+1$) &$k_{x- y},k_{x+ y},k_{z} $& $-$\\

&{\color{red}P}&$D_{2d}^{1}\times Z_{2}^{\tilde{T}}$& $2|4$ &($+1,-1,-1,+1$) &$k_{z},k_{x}^{2}-k_{y}^{2},k_{x}k_{y}$ &$-$  \\
\hline
142.566~\textrm{III}
&{\color{red}P}&$D_{2d}^{1}\times Z_{2}^{\tilde T}$& $2|4$ &($+1,-1,-1,+1$)&$k_{z},k_{x}^{2}-k_{y}^{2},k_{x}k_{y}$ &$-$\\
\hline

142.568~\textrm{III}&Z&$D_{2d}^{2}\times Z_{2}^{\tilde T}$& $4$ &($-1,-1,+1,-1$) &$[k_{x},k_{y}],k_{z}$& $-$\\
&X&$D_{2}^{2}\times Z_{2}^{\tilde T}$& $4$ &($-1,-1,-1,+1$) &$k_{x- y},k_{x+ y},k_{z}$& $-$\\
\hline
142.569~\textrm{III}&Z&$D_{4}^{1}\times Z_{2}^{\tilde T}$& $4$ &($-1,-1,+1,-1$) &$[k_{x},k_{y}],k_{z}$& $-$\\
\hline
\end{tabular}
\end{table*}

\begin{table*}[htbp]
(Extension of table \ref{tab: NodalDisp})\\
\centering
\begin{tabular}{ |c|ccc|c|c|c| }
\hline
Shubnikov s.g.&HSP& $G_{0}(\pmb k)$ &dim & l.c.&disp.&n.l.\\
\hline

201.21~\textrm{IV}&{\color{red}$\Gamma$} &$T_{h}\times Z_{2}^{T}$& $2|4_{}$ &($-1,-1,-1$) &$[k_{x}k_{y},k_{x}k_{z},k_{y}k_{z}]$ &$-$ \\
 &M&$D_{2h}^{1}\times Z_{2}^{T}$& $4$ &($-1,-1,-1,-1,-1,-1,-1$) &$k_{x},k_{y},k_{x}k_{y}k_{z},...$& $k_{z}$\\
&R&$T_{h}\times Z_{2}^{T}$& $4_{}$ &($-1,+1,-1$) &
$[k_{x}k_{y},k_{x}k_{z},k_{y}k_{z}]$& $k_{x},k_{y},k_{z}$\\
\hline

202.24~\textrm{III} &{\color{red}$\Gamma$} &$T\times Z_{2}^{\tilde{T}}$& $2|4$ &($-1,+1$) &$[k_{x},k_{y},k_{z}]$ &$-$ \\
\hline

202.25~\textrm{IV}&{\color{red}$\Gamma$} &$T_{h}\times Z_{2}^{T}$& $2|4_{}$ &($-1,-1,-1$) &$[k_{x}k_{y},k_{x}k_{z},k_{y}k_{z}]$ &$-$\\
\hline

203.28~\textrm{III} &{\color{red}$\Gamma$} &$T\times Z_{2}^{\tilde{T}}$& $2|4$ &($-1,+1$) &$[k_{x},k_{y},k_{z}]$ &$-$\\

&X&$D_{2}^{1}\times Z_{2}^{\tilde T}$&$4$ &($-1,-1,-1,+1$) &$k_{x},k_{y},k_{z}$& $-$\\
\hline
\end{tabular}
\end{table*}

\begin{table*}[htbp]
(Extension of table \ref{tab: NodalDisp})\\
\centering
\begin{tabular}{ |c|ccc|c|c|c| }
\hline
Shubnikov s.g.&HSP& $G_{0}(\pmb k)$ &dim & l.c.&disp.&n.l.\\
\hline

223.109~\textrm{IV}&{\color{red}$\Gamma$} &$O_{h}\times Z_{2}^{T}$& $2|4_{}$ &($-1,-1,-1,+1,+1$) &$[k_{x}k_{y},k_{x}k_{z},k_{y}k_{z}],$  & $-$\\
&&&&& $[2k_{z}^{2}-k_{x}^{2}-k_{y}^{2},k_{x}^{2}-k_{y}^{2}]$ & \\

&X&$D_{4h}^{2}\times Z_{2}^{T}$& $4$&($-1,+1,-1,+1,+1,+1,-1$) &
$k_{y},k_{x}k_{z},[k_{y}k_{x},k_{y}k_{z}]$& $k_{x},k_{z}$\\
 &M&$D_{4h}^{1}\times Z_{2}^{T}$& $4$&($-1,-1,-1,-1,+1,+1,+1$) &
 $k_{x}k_{y},[k_{z}k_{x},k_{z}k_{y}]$& $k_{x},k_{y},k_{z}$\\
 &R&$O_{h}\times Z_{2}^{T}$& $4_{1,2}$&($-1,+1,-1,+1,-1$) &
 $[k_{x}k_{y},k_{x}k_{z},k_{y}k_{z}]$& $k_{x},k_{y},k_{z}$\\
 &&&$4_{3}$&($-1,+1,-1,+1,-1$)& $[k_{x}k_{y},k_{x}k_{z},k_{y}k_{z}],k_{x}k_{y}k_{z},...$& $k_{x},k_{y},k_{z}$\\
\hline

224.112~\textrm{III}&{\color{red}$\Gamma$}&$T_{d}^{}\times Z_{2}^{\tilde T}$&$2|4$ &($-1,+1,+1$)&$[k_{x},k_{y},k_{z}]$& $-$\\
\hline

\end{tabular}
\end{table*}


\begin{table*}[htbp]
(Extension of table \ref{tab: NodalDisp})\\
\centering
\begin{tabular}{ |c|ccc|c|c|c| }
\hline
Shubnikov space group&HSP& $G_{0}(\pmb k)$ &dim & l.c.&disp.&n.l.\\
\hline

229.144~\textrm{III}&{\color{red}$\Gamma$}&$O\times Z_{2}^{\tilde T}$&$2|4$ &($-1,+1,+1$)&$[k_{x}k_{y},k_{x}k_{z},k_{y}k_{z}],$  & $-$\\
&&&&& $[2k_{z}^{2}-k_{x}^{2}-k_{y}^{2},k_{x}^{2}-k_{y}^{2}]$ & \\

&{\color{red}H}&$O\times Z_{2}^{\tilde T}$& $2|4$ &($-1,+1,+1$) &$[k_{x}k_{y},k_{x}k_{z},k_{y}k_{z}],$  & $-$\\
&&&&& $[2k_{z}^{2}-k_{x}^{2}-k_{y}^{2},k_{x}^{2}-k_{y}^{2}]$ & \\

&{\color{red}P}&$T\times Z_{2}^{\tilde{T}}$& $2|4$ &($-1,+1$) &$[k_{x},k_{y},k_{z}]$ &$-$ \\
\hline

230.147~\textrm{III}&{\color{red}$\Gamma$}&$T_{d}^{}\times Z_{2}^{\tilde T}$&$2|4$ &($-1,+1,+1$)&$[k_{x},k_{y},k_{z}]$& $-$\\
&{\color{red}H}&$T_{d}\times Z_{2}^{\tilde{T}}$& $2|4$ &($-1,+1,+1$) &$[k_{x},k_{y},k_{z}]$& $-$\\
&{\color{red}P}&$T_{d}\times Z_{2}^{\tilde{T}}$& $2|6$ &($-1,-1,+1$) &$[k_{x},k_{y},k_{z}]$& $-$\\
\hline

230.149~\textrm{III}&{\color{red}$\Gamma$}&$O\times Z_{2}^{\tilde T}$&$2|4$ &($-1,+1,+1$)&$[k_{x}k_{y},k_{x}k_{z},k_{y}k_{z}],$  & $-$\\
&&&&& $[2k_{z}^{2}-k_{x}^{2}-k_{y}^{2},k_{x}^{2}-k_{y}^{2}]$ & \\

&N&$D_{2}^{2}\times Z_{2}^{\tilde T}$& $4$ &($-1,-1,-1,+1$) &$k_{x- y},k_{x+ y},k_{z}$& $-$\\
&H&$O\times Z_{2}^{\tilde T}$& $4|4$ &($-1,+1,-1$) &$[k_{x},k_{y},k_{z}]$& $-$\\
&&&$4|8$ &($-1,+1,-1$) &$[k_{x},k_{y},k_{z}]$& $-$\\
&{\color{red}P}&$T\times Z_{2}^{\tilde{T}}$& $2|6$ &($-1,-1$) &$[k_{x},k_{y},k_{z}]$ &$-$ \\
\hline
\end{tabular}
\end{table*}

\appendix

\section{Projective Reps of anti-unitary groups}\label{regularProj}

{\bf Projective Reps and the factor systems}. As discussed in the main text, the little co-groups have the structure $G=H\times Z_{2}^{\mathbb T}$ with $Z_{2}^{\mathbb T}=\{E,\mathbb T\},\ \mathbb T^2=E$, where the physical meaning of $\mathbb T$ will be specified later. Supposing $g$ is a group element of $G$, then it is represented by $M(g)$ if $g$ is a unitary element and represented by $M(g)K$ if $g$  is anti-unitary, where $K$ is the complex-conjugate operator satisfying $KU=U^*K$ with $U$ an arbitrary matrix and $U^*$ its complex conjugation.

The multiplication of (projective) Reps of $g_1, g_2$ depends on if they are unitary or anti-unitary. If we define $s(g)$
\[
s(g)=\left\{
\begin{aligned}
&1,& &{\ \rm if} \ g \ {\rm is\ unitary,\ \ \ }   \\
&-1,& &{\ \rm if}\ g \ {\rm is\ antiunitary,\ \ \ }
\end{aligned}
\right.
\]
and the corresponding operator $K_{s(g)}$
\[
K_{s(g)}=\left\{
\begin{aligned}
&I,& &{\ \rm if\ } s(g)=1,\  \  \\
&K,& &{\ \rm if\ }s(g)=-1,
\end{aligned}
\right.
\]
then we have the multiplication rule of a projective Rep,
\[
M(g_1)K_{s(g_1)}M(g_2)K_{s(g_2)} = M(g_1g_2)e^{i\theta_2(g_1,g_2)}K_{s(g_1g_2)},
\]
where the $U(1)$ phase factor $\omega_2(g_1,g_2) \equiv e^{i\theta_2(g_1,g_2)}$ is a function of two group variables and is called the {\it factor system}. If $\omega_2(g_1,g_2) = 1$ for any $g_1,g_2\in G$, then above projective Rep becomes a linear Rep.

Substituting above results into the associativity relation of the sequence of operations $g_1\times g_2\times g_3$, we can obtain
\Beq
&&M(g_1)K_{s(g_1)}M(g_2)K_{s(g_2)}M(g_3)K_{s(g_3)} \ \\
&=& M(g_1g_2g_3)\omega_2(g_1,g_2)\omega_2(g_1g_2,g_3) K_{s(g_1g_2g_3)}\\
&=& M(g_1g_2g_3)\omega_2(g_1,g_2g_3)\omega_2^{s(g_1)}(g_2,g_3)K_{s(g_1g_2g_3)},
\Eeq
namely,
\beq\label{2cocyl}
\omega_2(g_1,g_2)\omega_2(g_1g_2,g_3) = \omega_2^{s(g_1)}(g_2,g_3)\omega_2(g_1,g_2g_3).
\eeq
Eq.(\ref{2cocyl}) is the general relation that the factor systems of any finite group (no matter unitary or anti-unitary) should satisfy.
If we introduce a gauge transformation $M'(g)K_{s(g)}=M(g)\Omega_1(g)K_{s(g)}$, where the phase factor $\Omega_1(g)=e^{i\theta_1(g)}$ depends on a single group variable, then the factor system changes into
\beq\label{gaugeomega2}
\omega'_2(g_1,g_2)=\omega_2(g_1,g_2)\Omega_2(g_1,g_2),
\eeq
with
\beq\label{2cob2}
\Omega_2(g_1,g_2) = {\Omega_1(g_1)\Omega_1^{s(g_1)}(g_2)\over \Omega_1(g_1g_2)}.
\eeq
The equivalent relations (\ref{gaugeomega2}) and (\ref{2cob2}) define the equivalent classes of the solutions of (\ref{2cocyl}). The number of equivalent classes for a finite group is usually finite.

{\bf The  2-cocycles and the 2$^{\rm nd}$ group cohomology.}  The factor systems (\ref{2cocyl}) of projective Reps of group $G$ are also called cocycles. The equivalent classes of the cocycles 
form a group, $i.e.$ the group-cohomology group. The group cohomology \footnote{For an introduction to group cohomology, see wiki {\href{http://en.wikipedia.org/wiki/Group_cohomology}{http://en.wikipedia.org/wik/Group\_cohomology}} and
Romyar Sharifi,  {\href{http://citeseerx.ist.psu.edu/viewdoc/summary?doi=10.1.1.296.244}{AN INTRODUCTION TO GROUP COHOMOLOGY}}}
$\{{\rm Kernel}\ d/{\rm Image}\ d\}$ is defined by the coboundary operator $d$
\begin{eqnarray}
& & (d\omega_{n})(g_{1},\ldots,g_{n+1}) \nonumber\\
&& =[g_{1}\cdot\omega_{n}(g_{2},\ldots,g_{n+1})]\omega^{(-1)^{n+1}}_{n}(g_{1},\ldots,g_{n})\times  \nonumber\\
&&
\prod^{n}_{i=1}\omega^{(-1)^{i}}_{n}(g_{1},\ldots,g_{i-1},g_{i}g_{i+1},g_{i+2},\ldots,g_{n+1}).
\end{eqnarray}
where $g_1,...,g_{n+1}\in G$ and the variables $\omega_{n}(g_{1},\ldots,g_{n})$ take value in an Abelian coefficient group $\mathcal A$ [usually $\mathcal A$ is a subgroup of $U(1)$, in the present work $\mathcal A=U(1)$]. 
The set of variables $\omega_{n}(g_{1},\ldots,g_{n})$ is called a $n$-cochain. For anti-unitary groups the module $g\cdot$ is defined by
\begin{eqnarray*}\label{module}
g\cdot\omega_{n}(g_{1},\ldots,g_{n}) =\omega_{n}^{s(g)}(g_{1},\ldots,g_{n}).
\end{eqnarray*}

With this notation, Eq. (\ref{2cocyl}) can be rewritten as
\Beq
(d\omega_2)(g_1,g_2,g_3)=1,
\Eeq
the solutions of above equations are called 2-cocycles with $U(1)$-coefficient. Similarly, Eq. (\ref{2cob2}) can be rewritten as
\Beq
\Omega_2(g_1,g_2)=(d\Omega_1)(g_1,g_2),
\Eeq
where $\Omega_1(g_1),\Omega_1(g_2)\in U(1)$ and $\Omega_2(g_1,g_2)$ are called 2-coboundaries. Two 2-cocycles $\omega_2'(g_1,g_2)$ and $\omega_2(g_1,g_2)$ are equivalent if they differ by a 2-coboundary, see Eq. (\ref{gaugeomega2}). The equivalent classes of the 2-cocycles $\omega_2(g_1,g_2)$ form the second group cohomology $\mathcal H^2(G, U(1))$.

Writing  $\omega_{2}(g_{1},g_{2})=e^{i\theta_{2}(g_{1},g_{2})}$, where $\theta_{2}(g_{1},g_{2})\in[0,2\pi)$, then the cocycle equations $(d\omega_{2})(g_{1},g_2,g_{3})=1$ can be written in terms of linear equations,
\begin{eqnarray}
&&s(g_{1})\theta_{2}(g_{2},g_{3})-\theta_{2}(g_{1}g_{2},g_{3})+\theta_{2}(g_{1},g_{2}g_{3})\nonumber\\
&&  -\theta_{2}(g_{1},g_{2})=0.
\label{2cocycle}
\end{eqnarray}
Similarly, if we write $\Omega_1(g_1)=e^{i\theta_1(g_1)}$ and $\Omega_2(g_1,g_2)=e^{i\Theta_2(g_1,g_2)}$, then the 2-coboundary (\ref{2cob2})
can be written as
\beq
\Theta_{2}(g_{1},g_{2})= s(g_{1})\theta_{1}(g_{2})-\theta_{1}(g_{1}g_{2})+\theta_{1}(g_{1}) .
  \label{2coboundary}
\end{eqnarray}
The equal sign in (\ref{2cocycle}) and (\ref{2coboundary}) means equal mod $2\pi$. From these linear equations, we can obtain the solution space of the cocycle equations, as well as the classes that the solutions belong to. The set of classes forms a finite Abelian group, which labels the classification of the projective Reps.

{\bf Invariants of projective Reps.}  The second group-cohomology group is generated by a certain number of {\it invariants}. The invariants are, by definition, invariant under the gauge transformation (\ref{gaugeomega2}) and (\ref{2cob2}). They are formed by independent functions of the cocycles $\omega_{2}(g_{1},g_{2})$.
 For instance, for the unitary group $D_2=Z_2\times Z_2=\{E,P\}\times \{E,Q\}$ with $P^2=E, Q^2=E$, the classification of 2-cocycles is
 $$\mathcal H^2(D_2, U(1))=\mathbb Z_2,$$ there  is only one independent invariant which is given by $\chi = {\omega_2(P,Q)\over \omega_2(Q,P)}$. Another example is the simplest anti-unitary group $Z_2^\mathbb{T}= \{E,\mathbb{T}\}$, which has $$\mathcal H^2(Z_2^\mathbb{T}, U(1))=\mathbb Z_2,$$
with the invariant $\chi_\mathbb{T}=\omega_2({\mathbb{T}},{\mathbb{T}})$.

 Generally, for any group $G$, if $\mathcal{H}^{2}(G,U(1))=\mathbb Z_{2}^{n}$, then the 2-cocycles of $G$ have $n$ invariants, each taking value $+1$ or $-1$. The invariants of several anti-unitary groups are given in Table \ref{tab:invrts}.

{\bf Regular projective Reps and irreducible projective Reps.} For a given factor system, we can easily construct the corresponding regular projective Rep (the regular Rep twisted by the 2-cocycles) using the group space as the Rep space. The group element $g$ is not only an operator $\hat{g}$, but also a basis $|g\rangle$.
The operator $\hat{g}_{1}$ acts on the basis $|g_{2}\rangle$ as the following,
\begin{equation}\label{regrep}
\hat{g}_{1}|g_{2}\rangle=e^{i\theta_2(g_{1},g_{2})}K_{s(g_1)}|g_{1}g_{2}\rangle,
\end{equation}
or in matrix form
\beq\label{gaction1}
\hat g_1=M(g_1)K_{s(g_1)},
\eeq
with matrix element
\begin{equation}\label{gmatrix}
M(g_{1})_{g,g_{2}}=\langle g|\hat g_1|g_2\rangle=e^{i\theta_2(g_{1},g_{2})}\delta_{g,g_{1}g_{2}}.
\end{equation}

For any group element $g_{3}$, we have
\begin{eqnarray*}
\hat{g}_{1}\hat{g}_{2}|g_{3}\rangle&=& \hat{g}_{1}\left[e^{i\theta_2(g_{2},g_{3})}K_{s(g_2)}|g_{2}g_{3}\rangle\right] \\
&=&e^{i\theta_2(g_{1},g_{2}g_{3})}K_{s(g_1)}\left[e^{i\theta_2(g_{2},g_{3})}K_{s(g_2)}|g_{1}g_{2}g_{3}\rangle\right] \\
&=&e^{i\theta_2(g_{1},g_{2}g_{3})}e^{is(g_1)\theta_2(g_{2},g_{3})}K_{s(g_1g_2)}|g_{1}g_{2}g_{3}\rangle,
\Eeq
and
\Beq
\widehat{g_{1}g_{2}}|g_{3}\rangle&=&e^{i\theta_2(g_{1}g_{2},g_{3})}K_{s(g_1g_2)}|g_{1}g_{2}g_{3}\rangle.
\end{eqnarray*}
Comparing with the $2$-cocycle equation (\ref{2cocycle}), it is easily obtained that $\hat g_1\hat g_2=e^{i\theta_2(g_1,g_2)} \widehat{g_1g_2}$.  In matrix form, this relation reads
\begin{eqnarray}\label{MMProj}
M(g_{1})K_{s(g_1)}M(g_{2})K_{s(g_2)}= e^{i\theta_2(g_{1},g_{2})}M(g_{1}g_{2})K_{s(g_1g_2)}.\nonumber\\
\end{eqnarray}
Eq.(\ref{MMProj}) indicates that $M(g)K_{s(g)}$ is indeed a projective Rep of the group $G$. For the trivial 2-cocycle where $e^{i\theta_2(g_{1},g_{2})}=1$ for all $g_1, g_2\in G$, $M(g)K_{s(g)}$ reduces to the regular linear Rep of $G$.

For a given group $G$ and factor system $\omega_2(g_1,g_2)$, all of the inequivalent irreducible projective Reps can be obtained by reducing the regular projective Rep using the eigenfunction method. The details of the method is provided in Ref. \cite{jyzxliu2018} and they will not be repeated here. The complete results of IPReps of relevant anti-unitary groups for given factor systems are provided in Tabs.\ref{tab:IPReps}$\sim$\ref{tab:OhZ2T}.

\section{Verification of cocycle Eq.(\ref{2cocy2}) for the factor system}\label{proofeq2}

For the phase factor
\begin{equation}\label{phasemag1}
\omega_2(g_1,g_2)=e^{-i\pmb K_{1}\cdot\pmb \tau_{2}},
\end{equation}
where reciprocal lattice vector $\pmb K_{1}=s(g_1)(g_1^{-1}\pmb k -\pmb k)$, $\pmb \tau_{2}$ is fractional translation associated with $g_{2}\in G_{0}(\pmb k)$.

For any  $g_{1},g_{2},g_{3}\in G_{0}(\pmb k)$,
\begin{eqnarray}
&& \omega_2(g_1,g_2 g_3)\omega_2^{s(g_1)}(g_2,g_3) \nonumber\\
&=&e^{-i\pmb K_{1}\cdot\pmb \tau_{23}}e^{-is(g_{1})\pmb K_{2}\cdot\pmb \tau_{3}}    \nonumber\\
&=&e^{-i\pmb K_{1}\cdot\pmb \tau_{2}}e^{-i\pmb K_{1}\cdot g_{2}\pmb \tau_{3}}
e^{-is(g_{1})\pmb K_{2}\cdot\pmb \tau_{3}}    \nonumber\\
&=&e^{-i\pmb K_{1}\cdot\pmb \tau_{2}}e^{-is(g_2)g_{2}^{-1}\pmb K_{1}\cdot \pmb \tau_{3}}
e^{-is(g_{1})\pmb K_{2}\cdot\pmb \tau_{3}}    \nonumber\\
&=&e^{-i\pmb K_{1}\cdot\pmb \tau_{2}}e^{-is(g_2)s(g_1)[(g_{1}g_{2})^{-1}\pmb k-g_{2}^{-1}\pmb k]
\cdot \pmb \tau_{3}} \nonumber\\
&&
\cdot e^{-is(g_{1})s(g_2)(g_2^{-1}\pmb k -\pmb k)\cdot\pmb \tau_{3}}    \nonumber\\
&=&e^{-i\pmb K_{1}\cdot\pmb \tau_{2}}
e^{-is(g_{1}g_{2})[(g_{1}g_{2})^{-1}\pmb k-\pmb k]\cdot\pmb \tau_{3}}\nonumber\\
&=&e^{-i\pmb K_{1}\cdot\pmb \tau_{2}}e^{-i\pmb K_{12}\cdot\pmb \tau_{3}} \nonumber\\
&=&\omega_2(g_1,g_2)\omega_2(g_{1}g_{2},g_{3}),
\end{eqnarray}
where we use $\pmb \tau_{23}=\pmb \tau_{2}+g_2 \pmb \tau_{3}+\pmb R_{23}$ with the lattice vector $\pmb R_{23}$,
$\pmb K_{12}=s(g_{1}g_{2})[(g_{1}g_{2})^{-1}\pmb k-\pmb k]$,
and $\pmb K_{1}\cdot g_{2}\pmb \tau_{3}=s(g_2)g_{2}^{-1}\pmb K_{1}\cdot \pmb \tau_{3}$.
It is obvious that the cocycle equation (\ref{2cocyl}) is satisfied and the phase factor (\ref{phasemag1}) is factor system of projective Rep of $G_{0}(\pmb k)$ at HSPs of BZ.

\section{Special groups in Table \ref{tab: NodalDisp}}

For $G_{0}(\pmb k)=C_{2h}^{1}\times Z_{2}^{T}$, the class ($-1,-1,-1,-1$) and the class ($-1,+1,-1,-1$) each has a $4$-dim IPRep. For the class ($-1,+1,-1,-1$), the perturbations that lift the degeneracy belong to $A_u$(1-fold), $B_u$(1-fold) or $B_g$(3-fold) Reps,  while for the class ($-1,-1,-1,-1$) the perturbations belong to $A_u$(3-fold), $B_u$(1-fold) or $B_g$(1-fold) Reps. Since $k_z$ belongs to the Rep $A_u$, the dispersion along $k_z$ direction is linear. The $B_g$ Reps includes the quadratic terms $k_xk_z$, $k_yk_z$ and higher order terms. The $B_u$ Rep includes $k_x, k_y, k_x^3, k_y^3, k_x^2k_y, k_xk_y^2$. From equation (\ref{amu}), $B_u$ occurs only once in both of the classes ($-1,-1,-1,-1$) and ($-1,+1,-1,-1$). So we need to consider a combination of all of the possible terms, namely,
\begin{equation}
H_{B_u} \!\!=\!\! ( a k_x\! +b k_y\! + c k_x^3\! + d k_y^3\! + e k_x^2k_y\! + f k_xk_y^2+\!\!...)\Gamma^{(B_u)},
\\
\end{equation}
where $a\sim f$ are non-universal constants.
Apart from one special direction of the ($k_{x},k_{y}$) plane, $H_{B_u}$
has only linear dispersion terms $ak_{x}+bk_{y}$. But in the special direction, the cubic dispersion can appear.
In the direction perpendicular to the special direction, the dispersion is linear, so
\beq
H_{B_u} \!\!&=&\!\! ( a' k_m\! +b' k_n^3\!)\Gamma^{(B_u)},
\eeq
where $a',b'$ are non-universal constants, and $\pmb m\parallel (a,b,0)^T$, $\pmb n\parallel (b,-a,0)^T$ with $\pmb n\perp\pmb m$. Therefore, the leading dispersion terms includes $k_m, k_n^3, k_z$ in orthonormal bases $[\pmb b_m, \pmb b_n, \pmb b_z]$.
The special direction depends on the detail of material.


For $G_{0}(\pmb k)=D_{6}^{}\times Z_{2}^{\tilde{T}}$, class ($-1,-1,+1,-1$) has three inequivalent $4$-dim IPReps, one has linear dispersion $k_{z}$ and cubic dispersions $k_{x}^{3}-3k_{x}k_{y}^{2},k_{y}^{3}-3k_{y}k_{x}^{2}$, the other two have linear dispersions $[k_{x},k_{y}]$ and $k_{z}$ . All the three $4$-dim Reps have no nodal lines.

For $G_{0}(\pmb k)=D_{3h}^{2}\times Z_{2}^{\tilde T}$, class ($-1,-1,-1,+1$) has two inequivalent $4$-dim IPReps, one has linear dispersion $k_{z}$, the other has linear dispersion $k_{z}$ and quadratic dispersion $[k_{z}k_{x},k_{z}k_{y}]$. Both the two $4$-dim Reps have nodal lines in the directions of $k_{y},k_{\sqrt{3}x \pm y}$.

For $G_{0}(\pmb k)=D_{3h}^{1}\times Z_{2}^{\tilde{T}}$, class ($-1,-1,-1,-1$) has two inequivalent $4$-dim IPReps, one has linear dispersion $k_{z}$, the other has linear dispersion $k_{z}$ and quadratic dispersion $[k_{z}k_{x},k_{z}k_{y}]$. Both the two $4$-dim Reps have nodal lines in the directions of $k_{x},k_{x\pm\sqrt{3}y}$.

For $G_{0}(\pmb k)=O_{h}^{}\times Z_{2}^{T}$, class ($-1,+1,-1,+1,-1$) has three inequivalent $4$-dim IPReps, one has quadratic dispersion $[k_{x}k_{y},k_{x}k_{z},k_{y}k_{z}]$ and cubic dispersion $k_{x}k_{y}k_{z}$, the other two have quadratic dispersion $[k_{x}k_{y},k_{x}k_{z},k_{y}k_{z}]$. All the three $4$-dim Reps have nodal lines in the directions of $k_{x},k_{y},k_{z}$.

For $G_{0}(\pmb k)=O_{h}^{}\times Z_{2}^{T}$, class ($-1,+1,-1,+1,+1$) has one $8$-dim irreducible Rep and two $4$-dim irreducible Reps. The $8$-dim Rep has linear dispersion $[k_{x},k_{y},k_{z}]$, but no nodal lines. The two $4$-dim Reps have cubic dispersion $k_{x}k_{y}k_{z}$ and nodal lines in the directions of $k_{x},k_{y},k_{z},k_{x\pm y},k_{x\pm z},k_{y\pm z}$.

For $G_{0}(\pmb k)=D_{3}^{3}\times Z_{2}^{\tilde{T}}$, class ($-1,-1$) has one $4$-dim IPRep, it has linear dispersions in the direction of $k_{x+y+z}$ and the plane $k_{x+y+z}^{\perp}$, which is perpendicular to $k_{x+y+z}$.

For $G_{0}(\pmb k)=D_{3d}^{3}\times Z_{2}^{T}$, class ($+1,+1,-1,+1$) has one $4$-dim IPRep, it has linear dispersion in the direction of $k_{x+y+z}$ and quadratic dispersion $(k_{x+y+z}^{\perp})^{2}$ in the plane perpendicular to $k_{x+y+z}$.


\begin{table*}[htbp]
\caption{Correspondence between some abstract groups $H\times Z_{2}^{\mathbb{T}}$ and concrete groups, where $\mathbb T$ is an abstract anti-unitary operation with $\mathbb T^2=E$, $\mathcal{I}={\rm space\ inversion}$, $T={\rm time\ reversal}$, $\mathcal{M}$ = mirror reflection plane, $Z_{2}^{\mathbb{T}}=\{E,\mathbb{T}\}$, $Z_{2}^{T}=\{E, T\}$, $Z_{2}^{\tilde T}=\{E,\tilde T\}$ with $\tilde T=\mathcal IT$. In the fourth column, we list all the classification labels of projective Rep. The invariants are interpreted as the following: $\chi_T\equiv\omega_{2}(T,T)$, $\chi_{\tilde T}\equiv\omega_{2}(\tilde T,\tilde T)$, $\chi_{TC_{2}}\equiv\omega_{2}(TC_{2},TC_{2})$, $\chi_{T\mathcal{M}}\equiv\omega_{2}(T\mathcal{M},T\mathcal{M})$.} \label{tab:invrts}
\centering
\begin{tabular}{ |c|c|c|c| }
\hline
Abstract group& Concrete group & Generators & Classification labels\\
\hline
$(Z_{2}\times Z_{2})\times Z_{2}^{\mathbb{T}}$
&$C_{2v}^{1}\times Z_{2}^{\tilde T}$& $P=C_{2x},Q=\mathcal{M}_{y},\mathbb T=\tilde T=\mathcal IT$ &
($\frac{\omega_{2}(\mathcal{M}_{y},C_{2x})}{\omega_{2}(C_{2x},\mathcal{M}_{y})}$,$\chi_{\tilde T}$,
$\chi_{T\mathcal{M}_{x}}$, $\chi_{TC_{2y}}$) \\

&$C_{2v}^{2}\times Z_{2}^{\tilde T}$& $P=\mathcal{M}_{x},Q=C_{2y},\mathbb T=\tilde T=\mathcal IT$ &
($\frac{\omega_{2}(\mathcal{M}_{x},C_{2y})}{\omega_{2}(C_{2y},\mathcal{M}_{x})}$,$\chi_{\tilde T}$,
$\chi_{TC_{2x}}$,$\chi_{T\mathcal{M}_{y}}$) \\

&$C_{2v}^{3}\times Z_{2}^{\tilde T}$& $P=\mathcal{M}_{x},Q=\mathcal{M}_{y},\mathbb T=\tilde T=\mathcal IT$ &
($\frac{\omega_{2}(\mathcal{M}_{x},\mathcal{M}_{y})}{\omega_{2}(\mathcal{M}_{y},\mathcal{M}_{x})}$,$\chi_{\tilde T}$,
$\chi_{TC_{2x}}$,$\chi_{TC_{2y}}$) \\



&$C_{2v}^{4}\times Z_{2}^{\tilde T}$& $P=C'_{21},Q=\mathcal{M}_{v1},\mathbb T=\tilde T=\mathcal IT$&
($\frac{\omega_{2}(\mathcal{M}_{v1},C'_{21})}{\omega_{2}(C'_{21},\mathcal{M}_{v1})}$,$\chi_{\tilde T}$,
$\chi_{T\mathcal{M}_{d1}}$, $\chi_{TC''_{21}}$) \\

&$C_{2v}^{5}\times Z_{2}^{\tilde T}$& $P=\mathcal{M}_{d1},Q=C''_{21},\mathbb T=\tilde T=\mathcal IT$&
($\frac{\omega_{2}(\mathcal{M}_{d1},C''_{21})}{\omega_{2}(C''_{21},\mathcal{M}_{d1})}$,$\chi_{\tilde T}$,
$\chi_{TC'_{21}}$,$\chi_{T\mathcal{M}_{v1}}$) \\
\hline

\end{tabular}
\end{table*}

\begin{table*}[htbp]
\caption{IPReps of abstract anti-unitary groups that appeared in the main text. We only list the Rep matrices of the generators.  The symbols $\sigma_{x,y,z}$ are the three Pauli matrices,\ $I$ is the $2 \times 2$ identity matrix,\  $\omega=e^{i\frac{2\pi}{3}}$, \ $\omega^{\frac{1}{2}}=e^{i\frac{\pi}{3}}$, l.c. denotes `label of classification', $K$ means the complex conjugate operator, $(*)$ labels the double valued Rep.} \label{tab:IPReps}
\centering
 \begin{tabular}{ |c|ccc|c| }
  \hline
Group & & Rep of generators & & label of classification\\
\hline

$(Z_2\times Z_2)\times Z_{2}^{\mathbb{T}}$ &$ \mathbb{T}$&$P$&$Q$&\\ 
\hline
& $K$&$1$ & $1$&($+1,+1,+1,+1$)\\
& $K$&$1$ & $-1$&\\
& $K$&$-1$ & $1$&\\
& $K$&$-1$ & $-1$&\\
\hline

&$\sigma_{x}K$&$-I$ & $i\sigma_{z}$&($+1,+1,+1,-1$) \\
&$\sigma_{x}K$&$I$ & $i\sigma_{z}$&\\
\end{tabular}
%
%

%
%
%

 $(Z_4 \times Z_2)\times Z_2^\mathbb{T}$&$ \mathbb{T}$&$P$&$Q$ &\\
\hline
&$K$&$1$&$1$ & $(+1,+1,+1,+1)$\\
&$K$&$1$&$-1$ &  \\
&$K$&$-1$&$1$ & \\
&$K$&$-1$&$-1$ & \\
&$\sigma_x K$&$i\sigma_z$&$-I$ & \\
&$\sigma_x K$&$i\sigma_z$&$I$ & \\
\hline
&$\sigma_x K$&
$\left(\begin{array}{cc}e^{i\frac{\pi}{4}}&0\\0&e^{-i\frac{\pi}{4}}\end{array}\right)$&
$-I$& $(+1,+1,+1,-1)$\\
\hline

\end{tabular}
\end{table*}

\begin{table*}[htbp]
\centering
 (Extension of Table \ref{tab:IPReps})\\
 \begin{tabular}{ |c|ccc|c|}
 \hline
 Group & &Rep of generators & & l.c.\\
 \hline
$(Z_4 \times Z_2)\times Z_2^\mathbb{T}$&$ \mathbb{T}$&$P$&$Q$ &\\
\hline

&$\sigma_y K$&
$\left(\begin{array}{cc}e^{i\frac{\pi}{4}}&0\\0&e^{-i\frac{\pi}{4}}\end{array}\right)$&
$i\sigma_z$& $(+1,-1,+1,-1)$\\
&$\sigma_y K$&
$\left(\begin{array}{cc}e^{i\frac{\pi}{4}}&0\\0&e^{-i\frac{\pi}{4}}\end{array}\right)$&
$-i\sigma_z$& \\
&$\sigma_y K$&
$-\left(\begin{array}{cc}e^{i\frac{\pi}{4}}&0\\0&e^{-i\frac{\pi}{4}}\end{array}\right)$&
$-i\sigma_z$& \\
&$\sigma_y K$&
$-\left(\begin{array}{cc}e^{i\frac{\pi}{4}}&0\\0&e^{-i\frac{\pi}{4}}\end{array}\right)$&
$i\sigma_z$& \\
\hline
&$\sigma_y K$&
$\left(\begin{array}{cc}e^{i\frac{\pi}{4}}&0\\0&e^{-i\frac{\pi}{4}}\end{array}\right)$&
$-I$&  $(+1,-1,-1,+1)$\\
&$\sigma_y K$&
$\left(\begin{array}{cc}e^{i\frac{\pi}{4}}&0\\0&e^{-i\frac{\pi}{4}}\end{array}\right)$&
$I$& $(*)$ \\
&$\sigma_y K$&
$-\left(\begin{array}{cc}e^{i\frac{\pi}{4}}&0\\0&e^{-i\frac{\pi}{4}}\end{array}\right)$&
$I$&  \\
&$\sigma_y K$&
$-\left(\begin{array}{cc}e^{i\frac{\pi}{4}}&0\\0&e^{-i\frac{\pi}{4}}\end{array}\right)$&
$-I$&  \\
\hline

%

&$\sigma_y K$&$I$&$I$ & $(+1,-1,-1,-1)$\\
&$\sigma_y K$&$I$&$-I$ & \\
&$\sigma_y K$&$-I$&$I$ & \\
&$\sigma_y K$&$-I$&$-I$ & \\
&$\sigma_y K$&$i\sigma_z$&$I$ & \\
&$\sigma_y K$&$i\sigma_z$&$-I$ & \\
\hline
&$IK$&$i\sigma_z$&$i\sigma_x$ &  $(-1,+1,+1,+1)$\\
&$\sigma_x K$&$\sigma_z$&$i\sigma_x$  & \\
\hline
\end{tabular}
\end{table*}

\begin{table*}[htbp]
\centering
 (Extension of Table \ref{tab:IPReps})\\
 \begin{tabular}{ |c|ccc|c|}
 \hline
 Group & &Rep of generators & & l.c.\\
 \hline
$(Z_4 \rtimes Z_2)\times Z_2^\mathbb{T}$&$ \mathbb{T}$&$P$&$Q$ &\\
\hline

&$\sigma_{y}K$&$I$ & $I$&($+1,-1,-1,-1$) \\%
&$\sigma_{y}K$&$I$ & $-I$&\\
&$\sigma_{y}K$&$-I$ & $I$&\\
&$\sigma_{y}K$&$-I$ & $-I$&\\
&$\sigma_{x}\otimes\sigma_{y}K$&$iI\otimes\sigma_{z}$ & $\sigma_{z}\otimes\sigma_{x}$&\\
\hline

&$\sigma_{x}K$&$-\left(\begin{array}{cc}e^{i\frac{\pi}{4}}&0\\0&e^{-i\frac{\pi}{4}}\end{array}\right)$ & $i\sigma_{x}$&($-1,+1,+1,+1$) \\%
&$\sigma_{x}K$&$\left(\begin{array}{cc}e^{i\frac{\pi}{4}}&0\\0&e^{-i\frac{\pi}{4}}\end{array}\right)$ & $i\sigma_{x}$&\\
\hline

&$I\otimes\sigma_{x}K$&$\sigma_{z}\otimes \left(\begin{array}{cc}e^{i\frac{\pi}{4}}&0\\0&e^{-i\frac{\pi}{4}}\end{array}\right)$& $\sigma_{x}\otimes\sigma_{y}$&($-1,+1,+1,-1$) \\
\hline

& $\sigma_{z}\otimes\sigma_{x}K$&$\sigma_{z}\otimes \left(\begin{array}{cc}e^{i\frac{\pi}{4}}&0\\0&e^{-i\frac{\pi}{4}}\end{array}\right)$ & $i\sigma_{x}\otimes\sigma_{y}$&($-1,+1,-1,+1$) \\%
\hline

& $\sigma_{x}\otimes\sigma_{x}K$&$I\otimes \left(\begin{array}{cc}e^{i\frac{\pi}{4}}&0\\0&e^{-i\frac{\pi}{4}}\end{array}\right)$ & $\sigma_{z}\otimes\sigma_{y}$&($-1,+1,-1,-1$)\\
& $\sigma_{x}\otimes\sigma_{x}K$&$-I\otimes \left(\begin{array}{cc}e^{i\frac{\pi}{4}}&0\\0&e^{-i\frac{\pi}{4}}\end{array}\right)$ & $-\sigma_{z}\otimes\sigma_{y}$&\\
\hline

%

&$\sigma_{y}K$&$\left(\begin{array}{cc}e^{i\frac{\pi}{4}}&0\\0&e^{-i\frac{\pi}{4}}\end{array}\right)$ & $-\sigma_{x}$&($-1,-1,+1,+1$) \\%
&$\sigma_{y}K$&$-\left(\begin{array}{cc}e^{i\frac{\pi}{4}}&0\\0&e^{-i\frac{\pi}{4}}\end{array}\right)$ & $\sigma_{x}$&
$(*)$\\
\hline

& $I\otimes\sigma_{y}K$&$\sigma_{z}\otimes \left(\begin{array}{cc}e^{i\frac{\pi}{4}}&0\\0&e^{-i\frac{\pi}{4}}\end{array}\right)$ & $i\sigma_{x}\otimes\sigma_{x}$&($-1,-1,+1,-1$)\\%
\hline

&$I\otimes\sigma_{y}K$&$\sigma_{z}\otimes \left(\begin{array}{cc}e^{i\frac{\pi}{4}}&0\\0&e^{-i\frac{\pi}{4}}\end{array}\right)$& $-\sigma_{y}\otimes\sigma_{y}$&($-1,-1,-1,+1$) \\%
\hline

&$\sigma_{y}\otimes \sigma_{x}K$ &
$I\otimes \left(\begin{array}{cc}e^{i\frac{\pi}{4}}&0\\0&e^{-i\frac{\pi}{4}}\end{array}\right)$ &
$iI \otimes \sigma_{y}$ & ($-1,-1,-1,-1$)\\
&$\sigma_{y}\otimes \sigma_{x}K$&
$-I\otimes \left(\begin{array}{cc}e^{i\frac{\pi}{4}}&0\\0&e^{-i\frac{\pi}{4}}\end{array}\right)$ &
$-iI \otimes \sigma_{y}$ & \\
\end{tabular}
%

$(Z_6 \times Z_2)\times Z_2^\mathbb{T}$&$\mathbb{T}$&$P$&$Q$ &\\
 \hline
&$K$&$1$&$1$&$(+1,+1,+1,+1)$ \\
&$K$&$1$&$-1$&\\
&$K$&$-1$&$1$&\\
&$K$&$-1$&$-1$&\\
&$\sigma_{x}K$&
$\left(\begin{array}{cc}
\omega&0\\0& \omega^2\end{array}\right)$&$I$&\\
&$\sigma_{x}K$&
$\left(\begin{array}{cc}
\omega&0\\0& \omega^2\end{array}\right)$&$-I$&\\
&$\sigma_{x}K$&
$-\left(\begin{array}{cc}
\omega&0\\0& \omega^2\end{array}\right)$&$-I$&\\
&$\sigma_{x}K$&
$-\left(\begin{array}{cc}
\omega&0\\0& \omega^2\end{array}\right)$&$I$&\\
\hline
\end{tabular}
\end{table*}

\begin{table*}[htbp]
\centering
 (Extension of Table \ref{tab:IPReps})\\
 \begin{tabular}{ |c|ccc|c|}
  \hline
 Group & &Rep of generators & & l.c.\\
 \hline
$(Z_{6}\times Z_{2})\times Z_2^\mathbb{T}$&$\mathbb{T}$&$P$&$Q$ &\\
\hline

&$\sigma_{x}K$&
$\left(\begin{array}{cc}e^{i\frac{\pi}{6}}&0\\0&e^{-i\frac{\pi}{6}}\end{array}\right)$&
$-I$& $(+1,+1,+1,-1)$\\
&$\sigma_{x}K$&$i\sigma_{z}$&$-I$&  \\
&$\sigma_{x}K$&
$\left(\begin{array}{cc}e^{i\frac{\pi}{6}}&0\\0&e^{-i\frac{\pi}{6}}\end{array}\right)$&
$I$&  \\
&$\sigma_{x}K$&$i\sigma_{z}$&$I$&  \\
&$\sigma_{x}K$&
$-\left(\begin{array}{cc}e^{i\frac{\pi}{6}}&0\\0&e^{-i\frac{\pi}{6}}\end{array}\right)$&
$I$&  \\
&$\sigma_{x}K$&
$-\left(\begin{array}{cc}e^{i\frac{\pi}{6}}&0\\0&e^{-i\frac{\pi}{6}}\end{array}\right)$&
$-I$&  \\
\hline

&$\sigma_{x}K$&
$-\left(\begin{array}{cc}e^{i\frac{\pi}{6}}&0\\0&e^{-i\frac{\pi}{6}}\end{array}\right)$&
$i\sigma_{z}$&$(+1,+1,-1,+1)$\\
&$\sigma_{x}K$&
$\left(\begin{array}{cc}e^{i\frac{\pi}{6}}&0\\0&e^{-i\frac{\pi}{6}}\end{array}\right)$&
$-i\sigma_{z}$& \\
&$\sigma_{x}K$&
$-\left(\begin{array}{cc}e^{i\frac{\pi}{6}}&0\\0&e^{-i\frac{\pi}{6}}\end{array}\right)$&
$-i\sigma_{z}$& \\
&$\sigma_{x}K$&$\left(\begin{array}{cc}e^{i\frac{\pi}{6}}&0\\0&e^{-i\frac{\pi}{6}}\end{array}\right)$&
$i\sigma_{z}$& \\
&$\sigma_{x}K$&$i\sigma_{z}$&$i\sigma_{z}$& \\
&$\sigma_{x}K$&$i\sigma_{z}$&$-i\sigma_{z}$& \\
\hline
&$\sigma_{x}K$&
$\left(\begin{array}{cc}
\omega&0\\0& \omega^2\end{array}\right)$&$-i\sigma_{z}$& $(+1,+1,-1,-1)$\\
&$\sigma_{x}K$&
$\left(\begin{array}{cc}
\omega&0\\0& \omega^2\end{array}\right)$&$i\sigma_{z}$&  \\
&$\sigma_{x}K$&
$-I$&$-i\sigma_{z}$&  \\
&$\sigma_{x}K$&
$I$&$-i\sigma_{z}$&  \\
&$\sigma_{x}K$&
$-\left(\begin{array}{cc}
\omega&0\\0& \omega^2\end{array}\right)$&$i\sigma_{z}$&  \\
&$\sigma_{x}K$&
$-\left(\begin{array}{cc}
\omega&0\\0& \omega^2\end{array}\right)$&$-i\sigma_{z}$&  \\
\hline

%

&$\sigma_{y}K$&$-I$&$-i\sigma_{z}$&  $(+1,-1,+1,+1)$\\
&$\sigma_{y}K$&
$\left(\begin{array}{cc}
\omega&0\\0& \omega^2\end{array}\right)$&$-i\sigma_{z}$&  \\
&$\sigma_{y}K$&
$\left(\begin{array}{cc}
\omega&0\\0& \omega^2\end{array}\right)$&$i\sigma_{z}$&  \\
&$\sigma_{y}K$&$I$&$-i\sigma_{z}$&  \\
&$\sigma_{y}K$&
$-\left(\begin{array}{cc}
\omega&0\\0& \omega^2\end{array}\right)$&$i\sigma_{z}$&  \\
&$\sigma_{y}K$&
$-\left(\begin{array}{cc}
\omega&0\\0& \omega^2\end{array}\right)$&$-i\sigma_{z}$&  \\
\hline
&$\sigma_{y}K$&$i\sigma_{z}$&$i\sigma_{z}$& $(+1,-1,+1,-1)$\\
&$\sigma_{y}K$&
$\left(\begin{array}{cc}e^{i\frac{\pi}{6}}&0\\0&e^{-i\frac{\pi}{6}}\end{array}\right)$&$-i\sigma_{z}$& \\
&$\sigma_{y}K$&
$\left(\begin{array}{cc}e^{i\frac{\pi}{6}}&0\\0&e^{-i\frac{\pi}{6}}\end{array}\right)$&$i\sigma_z$& \\
&$\sigma_{y}K$&$i\sigma_{z}$&$-i\sigma_z$& \\
&$\sigma_{y}K$&
$-\left(\begin{array}{cc}e^{i\frac{\pi}{6}}&0\\0&e^{-i\frac{\pi}{6}}\end{array}\right)$&$-i\sigma_z$& \\
&$\sigma_{y}K$&
$-\left(\begin{array}{cc}e^{i\frac{\pi}{6}}&0\\0&e^{-i\frac{\pi}{6}}\end{array}\right)$&$i\sigma_z$& \\
\hline
&$\sigma_{y}K$&
$\left(\begin{array}{cc}e^{i\frac{\pi}{6}}&0\\0&e^{-i\frac{\pi}{6}}\end{array}\right)$&
$-I$ & $(+1,-1,-1,+1)$\\
&$\sigma_{y}K$&
$-\left(\begin{array}{cc}e^{i\frac{\pi}{6}}&0\\0&e^{-i\frac{\pi}{6}}\end{array}\right)$&
$-I$ & $(*)$\\
&$\sigma_{y}K$&
$\left(\begin{array}{cc}e^{i\frac{\pi}{6}}&0\\0&e^{-i\frac{\pi}{6}}\end{array}\right)$&
$I$ & \\
&$\sigma_{y}K$&$i\sigma_{z}$&$-I$ & \\
&$\sigma_{y}K$&
$-\left(\begin{array}{cc}e^{i\frac{\pi}{6}}&0\\0&e^{-i\frac{\pi}{6}}\end{array}\right)$&
$I$ & \\
&$\sigma_{y}K$&$i\sigma_{z}$&$I$ & \\
\hline
\end{tabular}
\end{table*}

\begin{table*}[htbp]
\centering
 (Extension of Table \ref{tab:IPReps})\\
 \begin{tabular}{ |c|ccc|c|}
  \hline
 Group & &Rep of generators & & l.c.\\
 \hline
$(Z_{6}\times Z_{2})\times Z_2^\mathbb{T}$&$\mathbb{T}$&$P$&$Q$ &\\
\hline

&$\sigma_{y}K$&$-I$&$-I$&  $(+1,-1,-1,-1)$\\
&$\sigma_{y}K$&$-\left(\begin{array}{cc}
\omega&0\\0& \omega^2\end{array}\right)$&$I$&  \\
&$\sigma_{y}K$&$-I$&$I$&  \\
&$\sigma_{y}K$&$-\left(\begin{array}{cc}
\omega&0\\0& \omega^2\end{array}\right)$&$-I$&  \\
&$\sigma_{y}K$&$I$&$-I$&  \\
&$\sigma_{y}K$&$I$&$I$&  \\
\end{tabular}
%

$(Z_6 \rtimes Z_2)\times Z_{2}^{\mathbb{T}}$&$\mathbb{T}$&$P$&$Q$&\\
\hline
&$K$&$1$&$1$&$(+1,+1,+1,+1)$\\
&$K$&$1$&$-1$&\\
&$K$&$-1$&$1$&\\
&$K$&$-1$&$-1$&\\
&$\sigma_{x}K$&$\left(\begin{array}{cc}
\omega&0\\0& \omega^2\end{array}\right)$&
$\sigma_{x}$&\\
&$\sigma_{x}K$&$-\left(\begin{array}{cc}
\omega&0\\0& \omega^2\end{array}\right)$ &
 $\sigma_{x}$&\\
\hline

&$\sigma_{x}K$&$I$&$i\sigma_{z}$&($+1,+1,+1,-1$)\\%
&$\sigma_{x}K$&$-I$&$i\sigma_{z}$&\\
&$\sigma_{y}\otimes \sigma_{y}K$&
$I\otimes\left(\begin{array}{cc}
\omega&0\\0& \omega^2\end{array}\right)$&
$iI\otimes \sigma_{y}$&\\
&$\sigma_{y}\otimes \sigma_{y}K$&
$-I\otimes\left(\begin{array}{cc}
\omega&0\\0& \omega^2\end{array}\right)$&
$iI\otimes \sigma_{y}$&\\
\hline

&$\sigma_{x}K$&$i\sigma_{z}$&$I$&($+1,+1,-1,+1$)\\%
&$\sigma_{x}K$&$i\sigma_{z}$&$-I$&\\
& $\sigma_{z}\otimes\sigma_{x}K$&
$\sigma_{z}\otimes \left(\begin{array}{cc}e^{i\frac{\pi}{6}}&0\\0&e^{-i\frac{\pi}{6}}\end{array}\right)$ & $\sigma_{y}\otimes\sigma_{y}$&\\
\hline

&$\sigma_{x}K$&$i\sigma_{z}$&$i\sigma_{z}$&$(+1,+1,-1,-1)$\\
&$\sigma_{x}K$&$i\sigma_{z}$&$-i\sigma_{z}$&\\
& $\sigma_{z}\otimes\sigma_{x}K$&$\sigma_{z}\otimes \left(\begin{array}{cc}e^{i\frac{\pi}{6}}&0\\0&e^{-i\frac{\pi}{6}}\end{array}\right)$ & $i\sigma_{x}\otimes\sigma_{y}$&\\
\hline

&$\sigma_{y}K$&$i\sigma_{z}$&$i\sigma_{z}$&$(+1,-1,+1,+1)$\\
&$\sigma_{y}K$&$i\sigma_{z}$&$-i\sigma_{z}$&\\
& $I\otimes\sigma_{y}K$&$\sigma_{z}\otimes \left(\begin{array}{cc}e^{i\frac{\pi}{6}}&0\\0&e^{-i\frac{\pi}{6}}\end{array}\right)$ & $i\sigma_{x}\otimes\sigma_{x}$&\\
\hline
\end{tabular}
\end{table*}

\begin{table*}[htbp]
\centering
 (Extension of Table \ref{tab:IPReps})\\
 \begin{tabular}{ |c|ccc|c|}
  \hline
 Group & &Rep of generators & & l.c.\\
 \hline
$(Z_{6}\rtimes Z_{2})\times Z_2^\mathbb{T}$&$\mathbb{T}$&$P$&$Q$ &\\
\hline

&$\sigma_{y}K$&$i\sigma_{z}$&$I$&($+1,-1,+1,-1$)\\%
&$\sigma_{y}K$&$i\sigma_{z}$&$-I$&\\
&$I \otimes \sigma_{y}K$&
$\sigma_{z}\otimes \left(\begin{array}{cc}e^{i\frac{\pi}{6}}&0\\0&e^{-i\frac{\pi}{6}}\end{array}\right)$ &
$\sigma_{y}\otimes \sigma_{x}$&\\
\hline

&$\sigma_{y}K$&$I$&$i\sigma_{z}$&($+1,-1,-1,+1$)\\%
&$\sigma_{y}K$&$-I$&$i\sigma_{z}$&$D_{3d}\times Z_{2}^{T}\ \ (*)$\\
&$\sigma_{y}K$&
$\left(\begin{array}{cc}
\omega&0\\0& \omega^2\end{array}\right)$&
$i\sigma_{y}$&\\
\end{tabular}
%

&$\sigma_{y}\otimes \sigma_{x}K$&
$I\otimes\left(\begin{array}{cc}e^{i\frac{\pi}{6}}&0\\0&e^{-i\frac{\pi}{6}}\end{array}\right)$&
$iI\otimes\sigma_{y}$&($-1,-1,+1,-1$)\\%
&$\sigma_{y}\otimes \sigma_{x}K$&
$-I\otimes\left(\begin{array}{cc}e^{i\frac{\pi}{6}}&0\\0&e^{-i\frac{\pi}{6}}\end{array}\right)$&
$iI\otimes\sigma_{y}$&\\
&$\sigma_{y}\otimes \sigma_{x}K$&
$I\otimes i\sigma_{z}$&
$iI\otimes\sigma_{y}$&\\
\hline

&$I\otimes \sigma_{y}K$&$\sigma_{z}\otimes\left(\begin{array}{cc}
\omega&0\\0& \omega^2\end{array}\right)$
&$\sigma_{x}\otimes \sigma_{x}$&($-1,-1,-1,+1$) \\%
&$I\otimes \sigma_{y}K$&$\sigma_{z}\otimes I$&$\sigma_{x}\otimes \sigma_{x}$&\\
\hline

&$\sigma_{z}\otimes\sigma_{y}K$&$\sigma_{z}\otimes I$&$i\sigma_{x}\otimes\sigma_{x}$&($-1,-1,-1,-1$)\\%
&$\sigma_{z}\otimes\sigma_{y}K$&
$\sigma_{z}\otimes \left(\begin{array}{cc}
\omega&0\\0& \omega^2\end{array}\right)$&$i\sigma_{x}\otimes\sigma_{x}$&\\
\hline
\end{tabular}
\end{table*}

\begin{table*}[htbp]
 (Extension of Table \ref{tab:IPReps})\\
\centering
 \begin{tabular}{ |c|ccc|c| }
 \hline
 Group & &Rep of generators & & l.c.\\
 \hline
 $A_4\times Z_2^\mathbb{T}$&$\mathbb{T}$&$P$&$Q$&\\
\hline

&$K$&$1$&$1$&($+1,+1$)\\%
&$\sigma_{x}K$&
$\left(\begin{array}{cc}
\omega&0\\0& \omega^2\end{array}\right)$&
$\left(\begin{array}{cc}
\omega^{2}&0\\0& \omega\end{array}\right)$&\\
& $\left(\begin{array}{ccc}
1& 0&0 \\0 &0&1\\0&1&0\end{array}\right)K$ &
$\left(\begin{array}{ccc}1& 0&0 \\0 &\omega^2&0\\0&0&\omega\end{array}\right)$&
 $\frac{1}{3}\left(\begin{array}{ccc}-1& 2\omega^{\frac{1}{2}} &2\omega^{-\frac{1}{2}} \\2\omega^{\frac{1}{2}}  &\omega^{-\frac{1}{2}} &2\\2\omega^{-\frac{1}{2}} &2&\omega^{\frac{1}{2}} \end{array}\right)$&\\
\hline

&$\sigma_{y}\otimes \sigma_{y}K$&
$I\otimes \left(\begin{array}{cc}
\omega&0\\0& \omega^2\end{array}\right)$&
$-I\otimes \frac{\sqrt{3}}{3}\left(\begin{array}{cc}
e^{-i\frac{\pi}{6}} &\sqrt{2}i\\ \sqrt{2}i & e^{i\frac{\pi}{6}} \end{array}\right)$&($+1,-1$)\\%
&$\sigma_x \otimes IK$&
$\left(\begin{array}{cccc}
1&0&0&0\\
0&\omega^2&0&0\\
0&0&1&0\\
0&0&0&\omega\end{array}\right)$&
$\frac{\sqrt{3}}{3}\left(\begin{array}{cccc}
i&\sqrt{2}e^{-i\frac{\pi}{6}}&0&0\\
\sqrt{2}e^{-i\frac{\pi}{6}}&e^{i\frac{\pi}{6}}&0&0\\
0&0&-i&\sqrt{2}e^{i\frac{\pi}{6}}\\
0&0&\sqrt{2}e^{i\frac{\pi}{6}}&e^{-i\frac{\pi}{6}}\end{array}\right)$&
\\
\hline

&$\sigma_{y}K$&
$-\left(\begin{array}{cc}
\omega&0\\0& \omega^2\end{array}\right)$&
$\frac{\sqrt{3}}{3}\left(\begin{array}{cc}
e^{-i\frac{\pi}{6}} &\sqrt{2}i\\ \sqrt{2}i & e^{i\frac{\pi}{6}} \end{array}\right)$&
($-1,+1$)\\
\hline

\end{tabular}
\end{table*}

\begin{table*}[htbp]
\caption{The Reps of $O(T_{d})\times Z_{2}^{\tilde{T}}\simeq (A_{4}\rtimes Z_{2})\times Z_2^\mathbb{T}$. The symbols $\sigma_{x,y,z}$ are the three Pauli matrices,\ $I$ is the $2 \times 2$ identity matrix,\  $\omega=e^{i\frac{2\pi}{3}}$, \ $\omega^{\frac{1}{2}}=e^{i\frac{\pi}{3}}$, l.c. denotes `label of classification', $K$ stands for complex conjugation, $(*)$ labels the double valued Rep.} \label{tab:OZ2T}
\centering
\begin{tabular}{ |c|cc|c|}
 \hline
 $O(T_{d})\times Z_{2}^{\tilde{T}}$ &$\tilde T$& $C_{31}^{-}$& l.c.\\
 &$C_{33}^{+}$ &$C_{2f}(\mathcal{M}_{df})$ & \\
 \hline
&$K$  &$1$   & ($+1,+1,+1$) \\
 &$1$  &$1$
;&  \\
  &$K$ &$1$ &  \\
 & $1$ & $-1$
 ; &   \\
&$\sigma_{x}K$ &
$\left(\begin{array}{cc}
\omega&0\\0& \omega^2\end{array}\right)$& \\
&$\left(\begin{array}{cc}
\omega^{2}&0\\0& \omega\end{array}\right)$ &
$\sigma_{y}$
; &  \\
\hline
\end{tabular}
\end{table*}

{\footnotesize
\begin{table*}[htbp]
\centering
 (Extension of Table \ref{tab:OZ2T})\\
\begin{tabular}{ |c|cc|c|}
 \hline
 $O(T_{d})$ &$\tilde T$& $C_{31}^{-}$& l.c.\\
$\times Z_{2}^{\tilde{T}}$ &$C_{33}^{+}$ &$C_{2f}(\mathcal{M}_{df})$ & \\
 \hline

&$\sigma_{y}\otimes \sigma_{y}K$ &
$I\otimes\frac{\sqrt{2}}{2}\left(\begin{array}{cc}
e^{-i\frac{\pi}{4}}&i\\i& e^{i\frac{\pi}{4}}\end{array}\right)$&  ($+1,-1,-1$) \\
&$I\otimes\frac{\sqrt{2}}{2}\left(\begin{array}{cc}
e^{i\frac{\pi}{4}}&i\\i& e^{-i\frac{\pi}{4}}\end{array}\right)$  &
$iI\otimes\sigma_{y}$
; &  \\
&$\sigma_{y}\otimes \sigma_{y}K$ &
$I\otimes\frac{\sqrt{2}}{2}\left(\begin{array}{cc}
e^{-i\frac{\pi}{4}}&i\\i& e^{i\frac{\pi}{4}}\end{array}\right)$&  \\
&$I\otimes\frac{\sqrt{2}}{2}\left(\begin{array}{cc}
e^{i\frac{\pi}{4}}&i\\i& e^{-i\frac{\pi}{4}}\end{array}\right)$  &
$-iI\otimes\sigma_{y}$
; &  \\

&$(\sigma_{y}\otimes I)\otimes\sigma_{y}K$ &
$-I\otimes\frac{\sqrt{2}}{4}\left(\begin{array}{cccc}
e^{-i\frac{\pi}{4}} &-1&-\sqrt{3} &-\sqrt{3}e^{-i\frac{\pi}{4}}\\
 1 & e^{i\frac{\pi}{4}}&\sqrt{3}e^{i\frac{\pi}{4}}&-\sqrt{3}\\
 -\sqrt{3}&-\sqrt{3}e^{i\frac{\pi}{4}}&e^{i\frac{\pi}{4}}&-1\\
 \sqrt{3}e^{-i\frac{\pi}{4}}&-\sqrt{3}&1& e^{-i\frac{\pi}{4}}\end{array}\right)$ &  \\
&$-I\otimes\frac{\sqrt{2}}{4}\left(\begin{array}{cccc}
e^{i\frac{\pi}{4}} &-1&\sqrt{3} &\sqrt{3}e^{i\frac{\pi}{4}}\\
 1 & e^{-i\frac{\pi}{4}}&-\sqrt{3}e^{-i\frac{\pi}{4}}&\sqrt{3}\\
 \sqrt{3}&\sqrt{3}e^{-i\frac{\pi}{4}}&e^{-i\frac{\pi}{4}}&-1\\
 -\sqrt{3}e^{i\frac{\pi}{4}}&\sqrt{3}&1& e^{i\frac{\pi}{4}}\end{array}\right)$ &
$i(I\otimes I)\otimes\sigma_{x}$ & \\
\hline
\end{tabular}
\end{table*}

\begin{table*}[htbp]
\centering
 (Extension of Table \ref{tab:OZ2T})\\
%
\end{table*}
}


\begin{table*}[htbp]
\caption{The Reps of $T_{h}\times Z_{2}^{T}\simeq (A_{4}\times Z_{2})\times Z_2^\mathbb{T}$.  The symbols $\sigma_{x,y,z}$ are the three Pauli matrices,\ $I$ is the $2 \times 2$ identity matrix,\  $\omega=e^{i\frac{2\pi}{3}}$, \ $\omega^{\frac{1}{2}}=e^{i\frac{\pi}{3}}$, l.c. denotes `label of classification', $K$ stands for complex conjugation, $(*)$ labels the double valued Rep.} \label{tab:ThZ2T}
\centering
\begin{tabular}{ |c|cc|c|}
 \hline

 $T_{h}\times Z_{2}^{T}$ &$T$& $C_{31}^{-}$&l.c.\\
 &$C_{33}^{+}$ &$\mathcal{I}$ & \\
\hline
&$K$&$1$& $(+1,+1,+1)$ \\
&$1$&$1$
;&\\
&$K$&$1$& \\
&$1$&$-1$
;& \\
&$\sigma_{x}K$&
$\left(\begin{array}{cc}
\omega&0\\0& \omega^2\end{array}\right)$&\\
&$\left(\begin{array}{cc}
\omega^2&0\\0& \omega\end{array}\right)$&
$I$
;&\\
\hline

\end{tabular}
\end{table*}

\begin{table*}[htbp]
\centering
 (Extension of Table \ref{tab:ThZ2T})\\
\begin{tabular}{ |c|cc|c|}
 \hline
 $T_{h}\times Z_{2}^{T}$ &$T$& $C_{31}^{-}$& l.c.\\
 &$C_{33}^{+}$ &$\mathcal{I}$ & \\
 \hline
 &$\sigma_{y}K$&$-I$&$(+1,-1,-1)$\\ 
 &$-I$&$I$;&\\
 &&&\\
  &$\sigma_{y}K$&$-I$&\\
 &$-I$&$-I$;&\\
 &&&\\
\hline
\end{tabular}
\end{table*}

\begin{table*}[htbp]
\centering
 (Extension of Table \ref{tab:ThZ2T})\\
%
\end{table*}

%
\begin{table*}[htbp]
\caption{The Reps of $O_{h}\times Z_2^T\simeq[(A_{4}\rtimes Z_{2})\times Z_{2}]\times Z_{2}^{\mathbb{T}}$.  The symbols $\sigma_{x,y,z}$ are the three Pauli matrices,\ $I$ is the $2 \times 2$ identity matrix,\  $\omega=e^{i\frac{2\pi}{3}}$, \ $\omega^{\frac{1}{2}}=e^{i\frac{\pi}{3}}$, l.c. denotes `label of classification', $K$ stands for complex conjugation, $(*)$ labels the double valued Rep.}\label{tab:OhZ2T}
\centering
\begin{tabular}{ |ccc|}
 \hline
  $T$& $C_{31}^{-}$&l.c. \\
 $C_{33}^{+}$ &$C_{2f}$ &$\mathcal{I}$ \\
 \hline
$K$  &$1$  &  ($+1,+1,+1,+1,+1$) \\%
 $1$  &$1$ &$1$ ;  \\
  & & \\
  $K$ &$1$ &  \\
  $1$ & $1$ &$-1$;    \\
  & & \\
\hline
\end{tabular}
\end{table*}
\begin{table*}[htbp]
(Extension of table \ref{tab:OhZ2T})\\
\begin{tabular}{ |ccc|}
 \hline
  $T$& $C_{31}^{-}$&l.c. \\
 $C_{33}^{+}$ &$C_{2f}$ &$\mathcal{I}$ \\
 \hline
$IK$&
$I$&
($+1,+1,+1,+1,-1$) \\%
$I$&
$\sigma_{x}$&
$\sigma_{z}$;\\
&&\\
$\sigma_{x}\otimes\sigma_{x}K$&
$I\otimes\left(\begin{array}{cc}\omega&0\\0& \omega^2\end{array}\right)$&
\\
$I\otimes\left(\begin{array}{cc}
\omega^{2}&0\\0& \omega\end{array}\right)$&
$I\otimes\sigma_{x}$&
$\sigma_{z}\otimes\sigma_{z}$;\\
&&\\
\hline
\end{tabular}
\end{table*}

\begin{table*}[htbp]
(Extension of table \ref{tab:OhZ2T})\\
%
\end{table*}

\begin{table*}[htbp]
(Extension of table \ref{tab:OhZ2T})\\
%
\end{table*}


\begin{table*}[htbp]
(Extension of table \ref{tab:OhZ2T})\\
%
\end{table*}


\begin{table*}[htbp]
(Extension of table \ref{tab:OhZ2T})\\
%
\end{table*}


\clearpage
\begin{thebibliography}{10}

\bibitem{xlsc11}X.L.Qi and S.C.Zhang, Rev. Mod. Phys. {\bf 83}, 1057 (2011).

\bibitem{mzkane10}M. Z. Hasan and C. L. Kane, Rev. Mod. Phys. {\bf 82}, 3045 (2010).

\bibitem{ravjzaa13}R.-J. Slager, A. Mesaros, V. Juricic and J. Zaanen, Nature Phys. {\bf 9}, 98 (2013).

\bibitem{ReadGreen2000} N. Read and D. Green, Phys.Rev.B {\bf 61} 10267 (2000).

\bibitem{xlsc09}X.L.Qi, T.L.Hughes, S.Raghu and S.C. Zhang, Phys. Rev. Lett. {\bf 102}, 187001 (2009).

\bibitem{negmagresist16}Li, Q., Kharzeev, D., Zhang, C. et al. Nature Phys.  {\bf 12}, 550 (2016).

\bibitem{negmagresist20}Xu, S., Bao, C., Guo, PJ. et al. Nat Commun. {\bf 11}, 2370 (2020).

\bibitem{planahallef17}S. Nandy, G. Sharma, A. Taraphder and S. Tewari, Phys. Rev. Lett. {\bf 119}, 176804 (2017).

\bibitem{neto09} A. H. Castro Neto, F. Guinea, N. M. R. Peres, K. S.
Novoselov, and A. K. Geim, Rev. of Mod. Phys.
{\bf 81}, 109 (2009).

\bibitem{CLKane2012} S. M. Young, S. Zaheer, J. C. Y. Teo, C. L. Kane, E. J. Mele  and A. M. Rappe,
Phys. Rev. Lett. {\bf 108}, 140405 (2012).

\bibitem{burkprl11}A. A. Burkov and L. Balents, Phys. Rev. Lett. {\bf 107}, 127205 (2011).

\bibitem{burkprb11}A. A. Burkov, M. D. Hook and L. Balents, Phys. Rev. B {\bf 84}, 235126 (2011).


\bibitem{BJYNagaosa14} B.-J. Yang and N. Nagaosa, Nature Commun. {\bf 5}, 4898 (2014).

\bibitem{ZFXD14}Z. Liu, B. Zhou, Y. Zhang, Z. Wang, H. Weng, D. Prabhakaran, S.-K. Mo, Z. Shen, Z. Fang,
X. Dai, et al., Science, {\bf 343}, 864 (2014).

\bibitem{PD14}Z. Liu, J. Jiang, B. Zhou, Z. Wang, Y. Zhang, H. Weng, D. Prabhakaran, S. Mo, H. Peng,
P. Dudin, et al., Nature Mater. {\bf 13}, 677 (2014).


\bibitem{weng15}H. Weng, Y. Liang, Q. Xu, R. Yu, Z. Fang, X. Dai and Y. Kawazoe, Phys. Rev. B {\bf 92}, 045108 (2015).

\bibitem{cf15}C. Fang, Y. Chen, H.-Y. Kee and L. Fu, Phys. Rev. B {\bf 92}, 081201(R) (2015).

\bibitem{cf16}C. Fang, H. Weng, X. Dai and Z. Fang, Chin. Phys. B {\bf 25}, 117106 (2016).

\bibitem{watan16}H. Watanabe, H. C. Po, M. P. Zaletel, and A. Vishwanath, Phys. Rev. Lett. {\bf 117}, 096404 (2016).

\bibitem{Benervig16Sci}Barry Bradlyn, Jennifer Cano, Zhijun Wang, M. G. Vergniory, C. Felser, R. J. Cava and B. Andrei Bernevig, Science, {\bf 353}, 6299 (2016).

\bibitem{bernevig17}B. Bradlyn, L. Elcoro, J. Cano, M. G. Vergniory, Z. Wang, C. Felser, M. I. Aroyo,
and B. A. Bernevig, Nature {\bf 547}, 298 (2017).

\bibitem{watan17}H. C. Po, A. Vishwanath, and H. Watanabe, Nat. Commun. {\bf 8}, 50 (2017).

\bibitem{kruth17}J. Kruthoff, J. Boer, J. Wezel, C. L. Kane and R. Slager, Phys. Rev. X {\bf 7}, 041069 (2017).

\bibitem{cf181}Z. Song, T. Zhang, Z. Fang and C. Fang, Nat Commun {\bf 9}, 3530 (2018).

\bibitem{cf182}Z. Song, T. Zhang and C. Fang, Phys. Rev. X {\bf 8}, 031069 (2018).

\bibitem{khalaf18}E. Khalaf, H. C. Po, A. Vishwanath  and H. Watanabe, Phys. Rev. X {\bf 8}, 031070 (2018).

\bibitem{armitage18}N.P. Armitage, E.J. Mele and A. Vishwanath, Rev. Mod. Phys. {\bf 90}, 015001 (2018).



\bibitem{cf19}T. Zhang, Y. Jiang, Z. Song, H. Huang, Y. He, Z. Fang, H. Weng, and C. Fang,
Nature {\bf 566}, 475 (2019).

\bibitem{bernevig19}M. G. Vergniory, L. Elcoro, C. Felser, N. Regnault, B. A. Bernevig, and Z. Wang,
Nature {\bf 566}, 480 (2019).

\bibitem{xgw19}F. Tang, H. C. Po, A. Vishwanath, and X. Wan, Nature {\bf 566}, 486 (2019).

\bibitem{cano21} J.Cano and B.Bradlyn, Annu. Rev. Condens.Matter Phys. {\bf 12},225(2021).

\bibitem{Bi3exp_21}Xingxia Cui, Yafei Li, Deping Guo, Pengjie Guo, Cancan Lou, Guangqiang Mei, Chen Lin, Shijing Tan, Zhengxin Liu, Kai Liu, Zhong-Yi Lu, Hrvoje Petek, Limin Cao, Wei Ji, and Min Feng, arXiv:2012.15220

\bibitem{Bi3thr_21}Deping Guo, Pengjie Guo, Shiliang Tan, Min Feng, Limin Cao, Zhengxin Liu, Kai Liu, Zhong-Yi Lu, and Wei Ji, arXiv:2012.15218

\bibitem{gangxu16}P. Tang, Q. Zhou, G. Xu, S.C. Zhang,
Nature Phys. {\bf 12}, 1100 (2016).

\bibitem{gangxu18} G. Hua, S. Nie, Z. Song, R. Yu, G. Xu and K. Yao,
Phys. Rev. B {\bf 98}, 201116(R) (2018).

\bibitem{watan18} H. Watanabe, H. C. Po and Ashvin Vishwanath, Sci. Adv. {\bf 4}, eaat8685 (2018).

\bibitem{mfvjur19}R. M. Geilhufe, F. Guinea and V. Juricic, Phys. Rev. B {\bf 99}, 020404(R) (2019).

\bibitem{cano19} J. Cano, B. Bradlyn and M. G. Vergniory, APL Materials {\bf 7}, 101125 (2019).

\bibitem{bernevig20} Y. Xu, L. Elcoro, Z. Song, B. J. Wieder, M. G. Vergniory, N. Regnault, Y. Chen, C. Felser and B. A.
Bernevig, Nature {\bf 586}, 702 (2020).

\bibitem{bernevigmtqc20} L. Elcoro, B. J. Wieder, Z. Song, Y. Xu, B. Bradlyn and B. A. Bernevig,
arXiv:2010.00598[cond-mat.mes-hall](2020).

\bibitem{rjslager20}  A. Bouhon, G. F. Lange and R.-Jan Slager,
Phys. Rev. B {\bf 103}, 245127 (2021).


\bibitem{MorHam}Morton Hamermesh, \textit{Group Theory and Its Application to Physical Problems}, Reading Mass: Addison-Wesley (1962).

\bibitem{BradleyCrack} C. J. Bradley and A. P. Cracknell, \textit{The Mathemattcal
Theory of Symmetry in Solids-Representation theory for point
groups and space groups}, Clarendon, Oxford (1972).

\bibitem{ChenJQRMP85}J.-Q. Chen, M.-J. Gao, and G.-Q. Ma, Rev. Mod. Phys.{\bf 57}, 211 (1985).

\bibitem{ChenJQ02} Jin Quan Chen, Jialun Ping and Fan Wang, \textit{Group Rep Theory For Physicists},  World Scientific Publishing Company (2002).

\bibitem{YLF20invariants} J Yang, Z.-X. Liu and C Fang, arXiv:2009.07864[cond-mat.mes-hall](2020).

\bibitem{jyzxliu2018} J. Yang and Z.-X. Liu , J. Phys. A: Math. Theor. {\bf 51} 025207 (2018).

\bibitem{YangYFL_CG}Zhen-Yuan Yang, Jian Yang, Chen Fang and Zheng-Xin Liu,
J. Phys. A: Math. Theor. {\bf 54} 265202 (2021).







\bibitem{u2pdin94}A. Purwanto, R.A. Robinson, L. Havela, V. Sechovsky, P. Svoboda, H. Nakotte, K. Prokes, F.R. de Boer, A. Seret, J.M. Winand, J. Rebizant and J.C. Spirlet, Phys. Rev. B {\bf 50}, 6792 (1994).

\bibitem{CeCo2P298}M. Reehuis, W. Jeitschko, G. Kotzyba, B. Zimmer, X. Hub et al., J. Allo. Comp.{\bf 266}, 54 (1998).

\bibitem{p2u66}Troc, R., Leciejewicz, J., Ciszewski, R. et al., Phys. Stat. Sol. {\bf 15}, 515 (1966).

\bibitem{NdZn87}H. Fujii, Y. Uwatoko, K. Motoya, Y. Ito and T. Okamoto, J. Magn. Mater. {\bf 63-64}, 114 (1987).



\bibitem{YFe4Ge201}P. S.-Papamantellos, J. R.-Carvajal, G. Andre, N. Duong, K. Buschow and P. Toledano,
          J. Magn. Mater. {\bf 236}, 14 (2001).

\bibitem{LuFe4Ge212}P.S.-Papamantellos, K.Buschow and J.R.-Carvajal, J. Magn. Mater. {\bf 324},3709 (2012).

 \bibitem{NpS95}G.H. Lander, P. Burlet, et al., Physica B, {\bf 215}, 7(1995).







\end{thebibliography}
\end{document}